\documentclass{article}

\usepackage{listings}
\usepackage{paralist}
\usepackage[shortlabels]{enumitem}
\usepackage{algpseudocode}
\usepackage{algorithm}
\usepackage{xcolor}
\usepackage{verbatim}
\usepackage{todonotes}
\usepackage{xspace}
\usepackage{multirow}
\usepackage{url}

\usepackage{lineno}

\usepackage{soul}

\newenvironment{change}{}{}
\newenvironment{change2}{}{}
\newenvironment{change3}{\color{black}}{\color{black}}

\definecolor{light-gray}{gray}{0.9}
\definecolor{dark-gray}{gray}{0.6}

\algnewcommand\algorithmicforeach{\textbf{for each}}
\algdef{S}[FOR]{ForEach}[1]{\algorithmicforeach\ #1\ \algorithmicdo}
\algnewcommand\algorithmicinput{\textbf{Input:}}
\algnewcommand\algorithmicoutput{\textbf{Output:}}
\algnewcommand\algorithmicdeps{\textbf{Dependencies:}}
\algnewcommand\algorithmicletblock{\textbf{Let:}}
\algnewcommand\Input{\item[\algorithmicinput]}
\algnewcommand\Output{\item[\algorithmicoutput]}
\algnewcommand\Deps{\item[\algorithmicdeps]}
\algnewcommand\LetBlock{\item[\algorithmicletblock]}

\algdef{SE}[DOWHILE]{Do}{doWhile}{\algorithmicdo}[1]{\algorithmicwhile\ #1}

\newcommand{\tool}{\textsc{Tecs}\xspace}
\newcommand{\klee}{\textsc{Klee}\xspace}
\newcommand{\kcg}{\textsc{Kcg}\xspace}
\newcommand{\scadelang}{\textsc{Scade}\xspace}
\newcommand{\scade}{\textsc{Scade}\xspace}
\newcommand{\lustre}{\textsc{Lustre}\xspace}
\newcommand{\esterel}{\textsc{Esterel}\xspace}
\newcommand{\saferc}{\textsc{SaferC}\xspace}
\newcommand{\misra}{\textsc{Misra}\xspace}

\title{Automatically Generating Test Cases for\\ Safety-Critical Software via Symbolic Execution}

\author{Elson Kurian, Daniela Briola, Pietro Braione, Giovanni Denaro}

\date{}

\begin{document}
\maketitle

\footnotesize
{\it DISCO, University of Milano-Bicocca, Viale Sarca 336, building U14, 20126, Milan, Italy}
{e.kurian@campus.unimib.it, \{daniela.briola, pietro.braione, giovanni.denaro\}@unimib.it}

\normalsize

\begin{abstract}
Automated test generation based on symbolic execution can be beneficial for systematically testing safety-critical software, to facilitate test engineers to pursue the strict testing requirements mandated by the certification standards, while controlling at the same time the costs of the testing process. At the same time,
the development of safety-critical software is often constrained with programming languages \begin{change3}
or coding conventions
\end{change3} that ban linguistic features which are believed to downgrade the safety of the programs, e.g., they do not allow dynamic memory allocation and \begin{change3}
variable-length arrays,
\end{change3} limit the way in which loops are used,  forbid  recursion,  and bound the complexity of control conditions.   
As a matter of facts, these linguistic features 
are also the main efficiency-blockers for the test generation approaches based on symbolic execution at the state of the art.

\begin{change3}
This paper contributes new evidence of the effectiveness of generating test cases with symbolic execution for a significant class of industrial safety critical-systems.
We specifically focus on \scadelang,\end{change3}
a largely adopted model-based development language for
safety-critical embedded software, and \begin{change3}we report on a case study in which we exploited symbolic execution to automatically generate test cases for a set of safety-critical programs developed in \scadelang.\end{change3} 
To this end,
we introduce a novel test generator that we developed in a recent industrial project on testing safety-critical railway software written in \scadelang,
and we report on our experience of using
this test generator for testing
a set of \scadelang programs that belong to the development of an on-board signaling unit for high-speed rail. The results provide empirically evidence that symbolic execution is indeed a viable approach for generating high-quality test suites for the safety-critical programs considered in our case study.

 \end{abstract}

{\bf Keyword} automated test generation, symbolic execution, safety-critical software


\section{Introduction} \label{sec:intro}

\emph{Safety-critical} software systems control life-critical and  mission-critical tasks in airplanes, trains, cars, nuclear power plants and patient monitoring tools. Since failures in these systems can have catastrophic consequences, they must be highly reliable~\cite{hatton1995safer}. For this reason, the certification authorities of the specific sectors usually 
impose strict standards on both the development and the quality control activities~\cite{do178c,en50128}, in order to ensure the highest possible confidence in the correct behavior of the developed systems.

In this context, \emph{automated} test case generation can play a crucial role for achieving the testing objectives mandated by the standards, while controlling at the same time the associated costs. While the problem of generating a set of test inputs for an arbitrary software program that cover a given target is undecidable, research aims at producing automatic tools that work well ``in practice'', i.e., in a sufficient number of common cases.
The techniques that are more used by current tools are based either on random testing~\cite{Duran:RTeval:TSE:1984,Chen:adaptiverandom:2010}, or on search-based testing~\cite{Briand2010}, or on symbolic execution~\cite{surveySymbExe2018, Cadar2013}. Random and search-based testing sample the input space of the target programs, either in a purely random fashion, or guided by the improvement of a \emph{fitness} function, whose value correlates with the coverage objectives to optimize~\cite{Duran:RTeval:TSE:1984,afl,Pacheco:Randoop:ICSE:2007,Tonella:EvolutionaryTesting:ISSTA:2004,Fraser:EvoSuite:ESECFSE:2011}.
On converse, symbolic execution~\cite{10.1109/TSE.1976.233817,king:symbolic:acm:1976,KLEE,sage,chipounov:s2e:TOCS:2012,Tillmann:pex:TAP:2008,Braione:JBSE:ESECFSE:2016,Braione:SUSHI:ISSTA:2017} systematically explores the execution paths of the program under test: it computes the execution conditions of the explored paths, and solves these execution conditions with the help of an automatic SMT (satisfiability-modulo-theories) solver as, e.g., Yices~\cite{Dutertre:Yices:CAV:2014}, STP~\cite{10.1007/978-3-540-73368-3_52} or Z3~\cite{DeMoura:Z3:TACAS:2008}. If a solution is found, this is a test input covering the path.

\begin{change2}
This paper investigates the viability of symbolic execution for automated test generation for safety-critical software.
The choice of using symbolic execution is motivated by the importance of fulfilling the relevant test objectives (e.g., the coverage targets required by the certification standards) while testing  safety-critical software. 
By exploring the program paths systematically, symbolic execution should be in principle able 
to generate at least a test case for every test objective that can be reached on at least an execution path, 
a goal that the random and search-based techniques cannot generally guarantee\end{change2}.
Nonetheless we are also aware of the major challenges 
of designing test generators based on symbolic execution, which result from common limitations of this technique:

\begin{enumerate}[(i)]
\item 
\begin{change}
Coping with the so called \emph{path explosion problem}:
\end{change}
Since the number of execution paths of a program grows exponentially with the amount of decision logic in the program, and is generally unbounded for programs that include recursive calls and loops governed with arbitrary conditions, symbolic execution seldom succeeds to analyze all execution paths in finite time.
On the contrary,
the systematic exploration approach often engages symbolic execution in a very fine-grained analysis of some specific parts of the program execution space, while leaving many other parts entirely untested. 
\item 
\begin{change}
Suitably handling non-numeric inputs, i.e., pointers or references to \emph{dynamically allocated, possibly recursive data structures}:
\end{change}
For the analysis to be precise, symbolic execution shall be able to discriminate the executions in which the references within the input objects in the heap could be either assigned to null-values, or be alias of each other, or yet correspond to distinct objects, respectively~\cite{visser:generalizaed:2003}. This further exacerbates the computational requirements for the analysis. The number of objects and object configurations to be discriminated could even be unbounded for inputs defined as recursive data structures. 
\item 
\begin{change}
Tolerating the limitations of SMT solvers in computing the solutions of \emph{complex path constraints}:
\end{change}
In symbolic execution,
failing to solve the execution conditions of a program path can depend on either the path being indeed \emph{infeasible}, i.e.,  not executable with any input, or the path constraints being too hard for the current SMT solver to be decided within the allowed time budget, or yet outside of the theories supported by the SMT solver.
In the latter cases, the solver is unable to either provide a solution or prove that a solution does not exist.
The inability of solving complex path constraints can result in missed test cases, or waste large portions of test budget in the analysis of execution paths that depend on unsatisfiable  conditions that the constraint solver failed to pinpoint.
\end{enumerate}

\begin{change2}
\begin{change3}This paper contributes new evidence in support of\end{change3} the research hypothesis that,
although the above issues hindering the practicality of symbolic execution may hold for many general-purpose programs, they \begin{change3}
have reduced impact\end{change3} for a significant class of industrial safety-critical systems, where symbolic execution can therefore work effectively.
\end{change2}
In fact, safety-critical software often relies on programming languages or conding standards that ban some linguistic features, based on the (empirically motivated) ground that those features are common causes of subtle failures. For example, 
 \begin{change2}
 one of the tenets of safety-critical software development is avoiding unbounded consumption of time or space resources at runtime, to cope respectively with divergence or crashes. For this reason languages for safety-critical software development like \saferc~\cite{hatton1995safer} and \scadelang\footnote{\url{https://www.ansys.com/it-it/products/embedded-software/ansys-scade-suite}} (used in the avionics and in the railway domains, respectively), or coding standards like \misra\footnote{\url{https://www.misra.org.uk/Publications/tabid/57/Default.aspx#label-dvg}} (required in the automotive industry) restrict what the programmers are allowed to do. Relevant restrictions include: forbidding programmers from allocating memory dynamically, instead requiring all the memory to be allocated by local or global variables with predictable size; statically bounding the maximum number of iterations of loops; and avoiding recursion.\end{change2} \begin{change3}Some consequences of this regime are that in such applications the total number of execution paths is finite, every execution path has a finite depth, and many programming constructs that yield an explosion in the size of the execution state space are not used. 
 \end{change3}

In particular, \begin{change3}this paper reports on a case study\end{change3} drawing on our experience with a recent project aimed to develop an on-board signaling unit for high-speed rail, following the ERTMS\footnote{\url{www.era.europa.eu/activities/european-rail-traffic-management-system-ertms}} standard specification, in which we have been recently involved with an industrial partner. This on-board unit is an embedded safety-critical component that shall handle signals from several track-side devices, e.g., transponders deployed along the railway and control units at the stations, and shall notify the driver or even activate the braking devices of the train under some danger conditions. It
is currently being implemented with \scadelang, a  system  modelling  language  and  a  model-based  development environment for embedded software  largely adopted in industry\footnote{\begin{change2}
Ansys, the company that commercializes \scadelang and the supporting \scade Suite model-based design environment, reports uses of \scadelang at Subaru for automotive applications, and for many other  safety-critical, embedded applications, including, avionics and flight control, autonomous vehicles and gas turbines. [\url{www.ansys.com/products/embedded-software/ansys-scade-suite}]
\end{change2}}~\cite{scadeExample1,scadeExample2,PetitDoche15,karg16,ScadeAdaptive,SCADEIntro,SCADEinBook}
and certified according to the CENELEC norms~\cite{CENELEC}.
As we discuss in more detail in Section~\ref{sec:background}, \scadelang allows to specify models with a formalism based on finite state machines,
that forbids constructs like dynamic memory allocation, 
variable-length arrays, non-statically-in-bound accesses to arrays, pointer arithmetic, recursion and unbounded loops. Thanks to these restrictions, \scadelang models can be automatically translated to equivalent C programs that guarantee the certification standards~\cite{scadeWhite,ScadeAvionics} required by ERA, the European Union Agency for Railway\footnote{\url{www.era.europa.eu}}. 

\begin{change3}
In the reported case study, we explored whether and to which extent the programming  constraints on which the safety-critical software developed in \scadelang depends 
enable the exploitation of 
symbolic execution for effective automated testing of such programs.\end{change3}
In detail, this paper makes the following contributions:
\begin{enumerate}[(i)]
\begin{change2}
\item We introduce
an original test generator for \scadelang programs based on 
symbolic execution.
We refer to this test generator as \tool (Test Engine for Critical Software). \tool builds on the symbolic executor \klee~\cite{KLEE} to render an efficient symbolic analysis of the C programs that the \scade environment compiles out of the original \scadelang models.

The originality of \tool is tightly related to the 
goal of our case study, in that \tool  makes several distinctive design choices that explicitly exploit the programming constraints guaranteed for programs in \scadelang. 
First, \tool exploits 
symbolic execution to systematically analyze the state machine model that the \scadelang program represents. To this end, \tool steers the symbolic executor \klee through multiple analysis passes of the C program that corresponds to the transition function of the state machine model.  In its algorithm, \tool relies on the knowledge that the \scade translator produces C programs with finite execution paths, thanks to the avoidance of unbounded loops and recursive calls, which guarantees the termination of 
each analysis pass. 
Second, \tool exploits the knowledge that all data structures are statically allocated and not recursive, and the size of all arrays is statically specified, which implies that all input data structures are always made of a finite set of statically identifiable distinct fields. Thus, \tool
initializes the input data structures at the beginning of symbolic execution by assigning symbolic values to all fields at any nesting level (including the items in all array-typed fields).
In this way, \tool induces a specialized, efficient symbolic execution method that shall not cope with discriminating the possible ways of initializing the input data structures and their internal references during the analysis.

\item We 
report new empirical data that show that \tool successfully computed test cases that both achieve high model coverage 
of a set of \scadelang programs developed by our industrial partner, 
\begin{change3}and revealed (once enriched with suitable assertion-style test oracles)  subtle, previously unknown faults for some considered programs. \end{change3}
In this way, our case study 
provides supporting evidence of both the effectiveness of \tool and 
\begin{change3}
the suitability of symbolic execution for generating test cases for the considered class of safety critical programs.
\end{change3}
\end{change2}

\item \begin{change2}
Furthermore, we report on our experience  with using the tool AFL~\cite{afl}, a test generator that is very popular for security vulnerability testing, as a possible replacement of \klee in our tool. AFL is
based on random and search-based input selection heuristics. The results clearly indicate the weaknesses of the random selection approach, which missed many test objectives,
further underscoring the beneficial impacts of a systematic exploration of the program state space as in our approach. \end{change2}

\item Yet, for some considered programs, we were able to compare the test cases that \tool automatically produced with the ones that were already manually designed by the developers. The comparison revealed interesting complementarities, thus confirming the usefulness and the effectiveness of our test generator, and  further supporting the
exploitability of symbolic execution to generate test cases for \scadelang models.  
\end{enumerate}

This paper is organized as follows. Section~\ref{sec:background} surveys the main characteristics of the \scadelang programming language, elaborates on the language restrictions that enable our test generation approach, and introduces a sample \scadelang program that we use as working example in the paper. Section~\ref{sec:generation} details the design of the test generator \tool, focusing in particular on the design choices by which \tool exploits the programming constraints that derive from \scadelang. Section~\ref{sec:caseStudy} reports on the case study in which we used \tool to generate test cases for a set of programs that belong to the on-board train unit developed by our industrial partner. Finally, Section~\ref{sec:related} surveys the related work in the field, and Section~\ref{sec:conclusions} outlines our conclusions and plans for future work on the topics of this paper.

 \section{Safety-Critical Development with \scadelang}
\label{sec:background}
In this section, we survey the main characteristics of the \scadelang programming language, motivate our research hypothesis on the exploitability of symbolic execution to generate test cases for programs in \scade,
and introduce a sample \scade program that we use as working example in the subsequent sections of the paper.

\subsection{\scadelang and Characteristics of the \scade Programs}

\scadelang is a system modelling language that allows the design, implementation and verification of reliable embedded software systems. Ansys Inc.\ develops the language and commercializes the \scade Suite development environment, that allows to design embedded cyber-physical systems based on the \scadelang language, simulate their behaviour, and generate qualifiable/certifiable 
code from the models. \scadelang is customarily used to develop high-assurance and safety-critical embedded systems in a wide range of application domains as, e.g., avionics, automotive and railway.

The \scadelang modelling language belongs to the family of the synchronous languages, such as \lustre~\cite{10.1109/5.97300} and \esterel~\cite{10.1016/0167-6423(92)90005-V}. Synchronous languages assume that all the communications and computations in the systems that their models represent are performed instantaneously. A \scadelang model is reactive, and structured as a collection of communicating finite-state machines, procedures and functions. 
Each state may have a hierarchical structure, similar in spirit to, but with richer semantics than, the Statecharts~\cite{statecharts} or UML state machine languages~\cite{manual:uml}. The computation of a \scadelang model is performed as a sequence of discrete steps referred to as \emph{execution cycles}. At each execution cycle the outputs and the next state of the model are calculated from the inputs and the current state. At the end of a cycle the execution of the model performs an instantaneous transition to the next state as it enters the next cycle. 
\begin{change3} A valid \scadelang model must enjoy the property of running each execution cycle in bounded space and time,
and \scadelang 
rejects models that are not deterministic or not deadlock-free.\end{change3}
\scadelang has both a textual and an equivalent graphical syntax, and the \scade Suite development environment allows to edit a model in either format.

Integrated in the \scade Suite development environment, the automatic code generator \kcg translates the \scadelang models to semantically equivalent programs in either the Ada or the C programming language.
In this paper we  consider the translation to C programs. The programs generated by \kcg are provably equivalent to the \scadelang models of which they are a translation. 
By virtue of the aforementioned properties of the \scadelang models, 
\begin{change3}
\kcg is able to translate them to C programs that also are deterministic, deadlock-free, and that run in bounded space and time. 
\end{change3}
Moreover, 
\begin{change3}in compliance with the \scadelang language, \kcg aims to ensure that\end{change3} the generated programs are both \emph{embeddable}, i.e., deployable in embedded, resource-constrained environments, and \emph{compliant} with the most demanding safety levels of certification standards as, e.g., DO-178C~\cite{do178c}, IEC 61508~\cite{iec61508}, EN 50128~\cite{en50128}, 
and ISO 26262~\cite{iso26262}. To this end, \kcg translates a \scadelang model to a program expressed in a suitable subset of the C programming language that does not contain programming constructs that are deemed ``intrinsically unsafe'' or unfriendly with resource-constrained environments. \begin{change3}A more precise characterization of the C language subset that \kcg uses as a target for the translation of \scadelang models follows:

\begin{itemize}
\item Its semantics is unambiguous and precise (e.g., no undefined behaviors);
\item It is ISO C18 compliant;
\item It conforms to the \misra C 2012 coding standard rules;
\item It has no dynamic memory allocation, all the memory is statically allocated (i.e., no use of heap memory or of variable-length arrays);
\item It has no recursive function calls;
\item All loops are statically bounded: their number of iterations is determined by constant values known at code generation time;
\item No expression has side effects (e.g., no use of pre and post increment operators);
\item The code is decomposed into elementary assignment statements to local variables (e.g., no assignment to formal parameters of functions);
\item There is no dynamic address calculation (``pointer arithmetic'' expressions), no variable aliasing (including the fact that arrays are always accessed by their declaration names via the array subscript operator), no pathological use of the array subscript operator;
\item All the array accesses are bounded within the respective array index ranges: the values of the indices of array accesses are determined by constant values known at code generation time.
\end{itemize}

The restrictions over the C language adopted by \kcg are motivated by the required compliance with the highest safety levels of the certification standards that the generated code must address. These standards discourage, or utterly forbid, the use of dynamic memory, aliasing, unbounded iteration and recursion, to ensure that the program always runs in bounded space and time. Furthermore, \kcg does not ever produce recursive data structures while translating \scade programs in C: indeed, the main purpose of recursive data structures is implementing unbounded containers, but since a well-formed \scadelang model always runs in bounded space there is no real need for its C translation to use unbounded containers. We remark that the nature of \scadelang models---their being deterministic, deadlock-free, and bounded in space and in time---is precisely what allows such a limited fragment of the C language to adequately express the full semantics of the \scadelang language. 
\end{change3}

\begin{change2}
In the target environment, the embedded 
software
must interact with the sensors and the actuators of the hardware platform.
In order to link the \scadelang programs to the hardware developers must implement suitable \emph{glue code}, i.e., peripheral drivers, 
\begin{change3}
interfacing the \kcg code generated from a \scadelang model and the external environment.
\end{change3}
We remark that the test generation problems that we 
consider
in this paper refer to \begin{change3}the inputs and the 
outputs of the \kcg-generated programs, i.e.,  the inputs and outputs of the \scadelang models,  regardless of the possible glue code that binds these inputs and outputs to sensors and actuators of the final system\end{change3}.
\end{change2}

\subsection{Working example}
We will use a simple \scadelang model to introduce the main concepts and terminology about \scadelang,  and to show how a \scadelang model is converted into C code: this will help the reader in better understanding how our approach described in Section \ref{sec:generation} works.

\begin{figure*}[t]
\centering
\includegraphics[width=\textwidth]{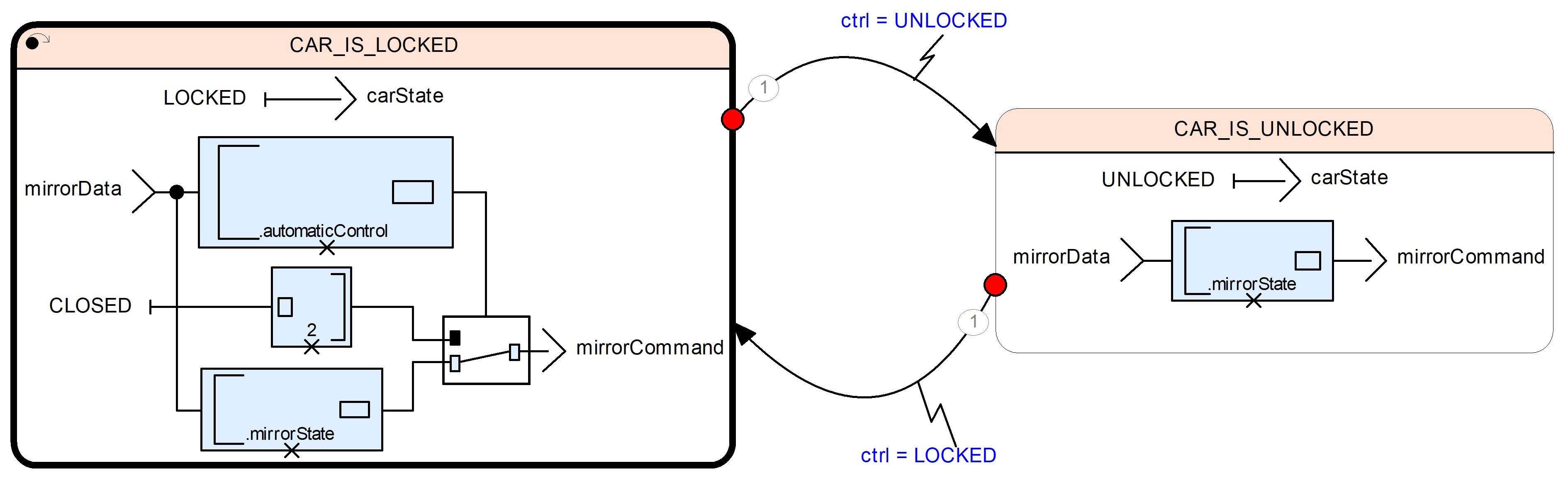} 

\caption{A sample \scadelang model for a car wing mirror controller} 
\label{fig:scade:program}
\end{figure*}

Figure~\ref{fig:scade:program} shows a \scadelang model that describes a simple controller for the wing mirrors of a car, for which it is possible to activate the behavior of closing the wing mirrors automatically when the car gets locked. The state machine has two states (the boxes in the left and right part of the figure, respectively) that represent whether the car is either locked or unlocked, respectively. The input signal $ctrl$ governs the possible transitions between these two states. The program starts in the state $CAR\_IS\_LOCKED$ (the state on the left of the figure) and then, if $ctrl$ gets set to $UNLOCKED$ the program changes state to $CAR\_IS\_UNLOCKED$ (the state on the right of the figure). Conversely, if $ctrl$ gets set to $LOCKED$ the program returns to  $CAR\_IS\_LOCKED$. The signal $ctrl$ can be thought as the input that the car receives from a remote controller. 

The program has three further inputs and three outputs. The three inputs are aggregated in the data structure $mirrorData$, which is
referred in both states of the \scade models in Figure~\ref{fig:scade:program}. 
The data structure $mirrorData$ consists of two fields: Field $mirrorData.automatic\-Control$ (dereferenced with the \scadelang operator represented as a rectangle in the top part of state $CAR\_\-IS\_\-LOCKED$) controls whether or not the automatic-closing behavior is currently active; Field $mirrorData.mirrorState$ (dereferenced in both program states) is an array of two items, each defining the latest state (either $OPEN$ or $CLOSED$) that the driver has set for either wing-mirror.
\begin{change2}The three outputs are $carState$, which records the current state of the car, and the two items of array $mirrorCommand$, which indicate the commands (either $OPEN$ or $CLOSED$) sent to the wing-mirrors.\end{change2}
The $carState$ 
is simply assigned as $LOCKED$ or $UNLOCKED$ in the two states of the program, respectively. \scadelang represents the assignment with an arrow that connects a value to the receiving variable, e.g., $LOCKED \rightarrow carState$ represents the assignment of the output 
$carState$ in the program state  $CAR\_\-IS\_\-LOCKED$. 

The main behavior of program is to define the commands sent to the wing mirrors when the control system is in each of the two program states, 
respectively. If the automatic-closing behavior is active, the wing mirrors shall close automatically upon locking the car. Otherwise, they shall just remain as they are. Upon unlocking the car, the wing mirrors shall always return as they were when the car got locked. The program encodes this behavior as follows. When the car gets locked (state $CAR\_IS\_LOCKED$)
the outputs $mirrorCommand$ are assigned with the if-then-else block represented as the white rectangle in the bottom-right part of state $CAR\_IS\_LOCKED$ in Figure~\ref{fig:scade:program}. The if-then-else block takes $mirrorData.automaticControl$ as condition (entering from the top of the block): 
If the automatic control is active, the outputs $mirrorCommand$ are both assigned as the constant $CLOSED$ (entering at the top-left corner of the block). Otherwise, if the automatic control is not active, they are assigned to the values in the array $mirrorData.mirrorState$ (entering at the bottom-left corner of the block). When the car gets unlocked (state $CAR\_IS\_UNLOCKED$) the outputs $mirrorCommand$ are always assigned the values of $mirrorData.mirrorState$. 

\begin{figure}[p]
\centering
\scriptsize

\begin{lstlisting}[language=C]
typedef struct {
  Lock ctrl;
  MirrorData mirrorData;
} inC_WingMirrorControl_CarControl;

typedef struct {
  MirrorStateArray mirrorCommand;
  Lock carState;
  SSM_ST_WingMirrorFSM WingMirrorFSM_state_nxt;
} outC_WingMirrorControl_CarControl;

typedef struct {
  kcg_bool automaticControl;
  MirrorStateArray mirrorState;
} MirrorData;

typedef MirrorState MirrorStateArray[2];

typedef enum {UNLOCKED, LOCKED} Lock;

typedef enum {OPEN, CLOSED} MirrorState;

void WingMirrorControl_CarControl(
  inC_WingMirrorControl_CarControl *inC,
  outC_WingMirrorControl_CarControl *outC);
{
  SSM_ST_WingMirrorFSM WingMirrorFSM_state_act;
  kcg_size idx;

  switch (outC->WingMirrorFSM_state_nxt) {
    case SSM_st_CAR_IS_UNLOCKED_WingMirrorFSM:
      if (inC->ctrl == LOCKED) {
        WingMirrorFSM_state_act = 
        SSM_st_CAR_IS_LOCKED_WingMirrorFSM;
      }
      else {
        WingMirrorFSM_state_act = 
        SSM_st_CAR_IS_UNLOCKED_WingMirrorFSM;
      }
      break;
    case ...
  }
  
  switch (WingMirrorFSM_state_act) {
    case SSM_st_CAR_IS_UNLOCKED_WingMirrorFSM:
      kcg_copy_WingMirrorArray(outC->mirrorCommand, 
      inC->mirrorData.mirrorState);
      outC->carState = UNLOCKED;
      outC->WingMirrorFSM_state_nxt = 
      SSM_st_CAR_IS_UNLOCKED_WingMirrorFSM;
      break;
    case ...
  }
}
\end{lstlisting}
\caption{Excerpt of the C program that \kcg generates for the \scadelang model in Figure~\ref{fig:scade:program}} 
\label{fig:c:program}
\end{figure}

Compiling the \scadelang program of Figure~\ref{fig:scade:program} with \kcg yields the C program excerpted in Figure~\ref{fig:c:program}. The program defines the entry function \texttt{WingMirror\-Control\_CarControl} (excerpted at the bottom of the figure) that encodes the behavior of the system. This function will be continuously executed at each execution cycle on the target board. As parameters, the program takes pointers to two data structures \texttt{inC} and \texttt{outC} of type \texttt{inC\_WingMirrorControl\_CarControl} and \texttt{outC\_WingMirrorControl\_CarControl}, respectively: \texttt{inC} wraps the inputs that the state machine receives at the beginning of each execution cycle, and
\texttt{outC} wraps the outputs of the state machine, along with a special field (\texttt{WingMirrorFSM\_state\_nxt}) that \kcg generates to encode the next state of the state machine after each execution cycle.
The top part of the code lists the type definitions for both \texttt{inC} and \texttt{outC} data structures, and their nested types. 

The body of the entry function consists of two switch statements executed in sequence. The first switch statement calculates the next state, and stores it in the temporary variable \texttt{WingMirrorFSM\_state\_act}. The second switch statement calculates the outputs, and assigns the fields of $\texttt{outC}$. For example, when the first switch statement computes the next state \texttt{SSM\_st\_CAR\_IS\_\-UNLOCKED\_\-WingMirrorFSM}, corresponding to the model state $CAR\_IS\_UNLOCKED$, the second switch statement assigns the outputs \texttt{outC->mirrorCommand} to the values of the inputs \texttt{inC->wingMirrorData.mirrorState}, the output \texttt{outC->carState} to  \texttt{UNLOCKED}, and the output \texttt{outC->WingMirrorFSM\_state\_nxt} to \texttt{SSM\_st\_CAR\_\-IS\_\-UNLOCKED\_WingMirrorFSM}.

 \section{Generating \scadelang Test Cases}
\label{sec:generation}

\begin{change3}
In this section, we introduce a
test generator to automatically generate unit-level test cases for embedded programs written in \scadelang.
Our test generator for \scadelang programs is built on top of the symbolic executor \klee, and
 explicitly relies on the 
C language restrictions that \kcg enforces (as we discussed in Section~\ref{sec:background}).
We designed the test generator with the aim of exploring whether and to which extent these restrictions identify a class of programs that by design mitigate many common sources of open issues for test generators based on symbolic execution.
In this section we describe the design of the test generator, while in the next section we report on the effectiveness of the test generator to derive test cases for a set of \scadelang programs implemented in a recent project in which we are participating along with an industrial partner.
\end{change3}

\begin{figure*}[t]
\includegraphics[width=0.95\textwidth]{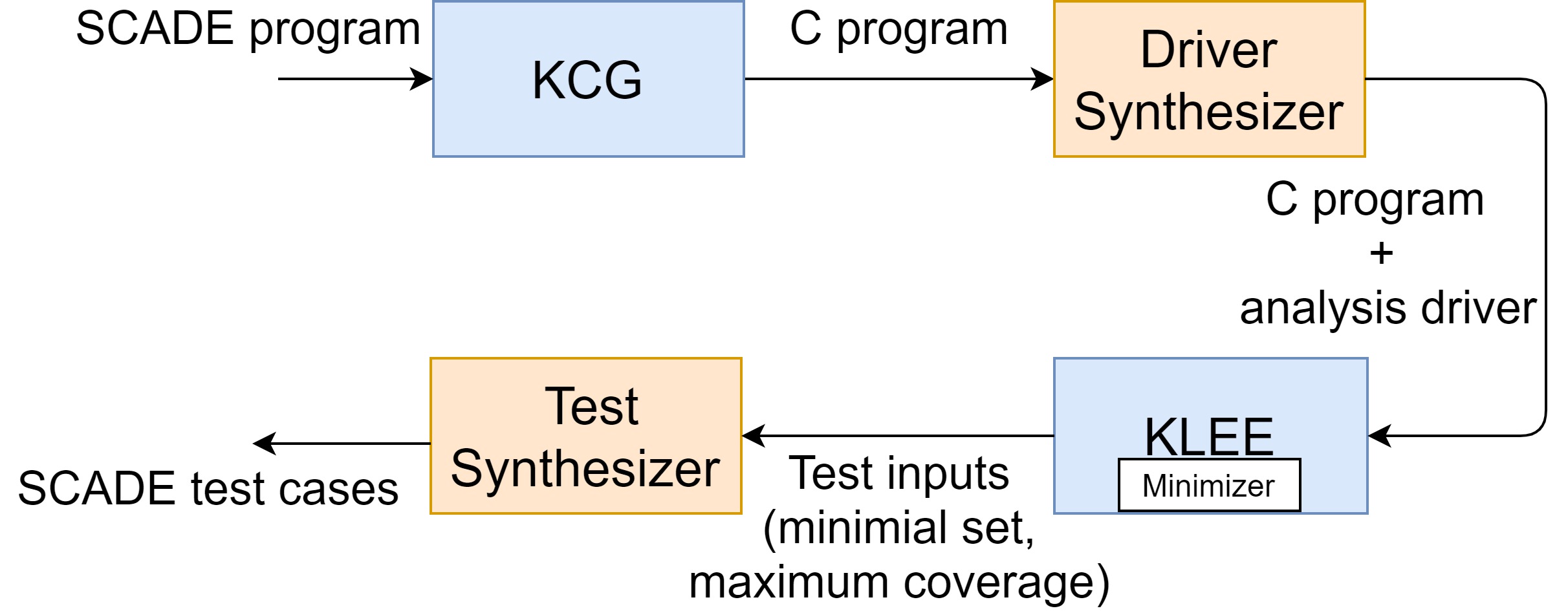} 
\caption{Workflow of \tool} 
\label{fig:tool}
\end{figure*}

Figure~\ref{fig:tool} shows the main components of our test generator, and the workflow that these components comprise. We refer to our test generator as \tool, the \emph{Test Engine for Critical Software}.
\tool relies on the \kcg compiler, which is part of the cross-compilation tool chain of the \scade Suite, to convert the \scadelang program under test into an equivalent program written in C.

Then, \tool includes a \emph{Driver synthesizer} that augments the obtained C program with an analysis driver written itself in C. 
\begin{change}
The analysis driver embodies the actual analysis algorithm that \tool uses to explore the state space of the program under test: It assigns the program inputs with symbolic values, and then calls the original program 
multiple times, aiming to trigger the possible transitions of the state machine model that the \scadelang program  represents. Thus, by executing the analysis driver with symbolic execution, \tool steers multiple analysis passes of the execution paths in the program, with each new pass depending on the (symbolically represented) results of the previous pass. As we explain in detail in Section~\ref{sec:driverSynthesizer}, the Driver synthesizer tailors the general analysis algorithm to the specific signature of the program under test.

To accomplish symbolic execution according to the analysis algorithm provided with the analysis driver, 
\end{change}
\tool relies on \klee, a well known state-of-the-art symbolic executor for programs in C~\cite{KLEE}. 
\klee  generates test inputs for each execution path that the analysis driver induces through the program
as follows. 
First, \klee  performs symbolic execution along each execution path to compute the associated \emph{path condition}, being the path condition a (quantifier free) logic formula that represents the conditions that the inputs shall satisfy for the program to execute along the given path. Then, \klee attempts to solve each path condition with the STP~\cite{GaneshD07} constraint solver. If a path condition has a solution, this is a set of concrete inputs for executing the corresponding execution path; Otherwise, if there is no solution, the path is \emph{infeasible}, i.e., no input can drive the execution of the program through it. 
\begin{change2} 
Yet, a further and unfortunate phenomenon is that some path conditions could be formulas that the STP constraint solver cannot solve within the allowed timeout (1  millisecond in our current experiments), and thus the test generation process might result in either or both missed test cases and wasted analysis time. As we already commented in the introduction section, this phenomenon is a common source of ineffectiveness for test generators based on symbolic execution, but also a phenomenon that we hypothesize to be rare for the safety-critical programs in \scadelang. Indeed, in the case study that we report in Section~\ref{sec:caseStudy} we did not experience any unsolved path condition. 
\end{change2} 

Next, \tool filters the execution paths explored during symbolic execution by computing the minimal set of execution paths that guarantee the maximum coverage of the relevant test objectives (Figure~\ref{fig:tool}, \emph{Minimizer}). 
\begin{change2}
In fact,  maintaining test suites that include a test case for each execution path is by far beyond the typical certification requirements, and  cannot be generally afforded by producers. 
The Minimizer aims to produce a test suite of manageable size, while avoiding to miss test objectives.
\end{change2}

\begin{change2}
As a limitation of our current prototype, 
\tool delegates
the task of the Minimizer to an internal algorithm of \klee, which can be optionally activated to limit the provided test inputs only to the execution paths that improve statement coverage. This is a sub-optimal minimization behavior, and indeed in the experiments that we report later in this paper we observed that some test objectives that are considered in the \scade Test tool (which refers to  modified condition/decision coverage, a finer criterion than statement coverage) were missed. We discuss the  results of the experiments in detail in Section~\ref{sec:caseStudy}. 
In future releases, we aim to improve our tool by providing a dedicated Minimizer, which is the reason why we illustrated the Minimizer as a logical component of the approach in Figure~\ref{fig:tool}.
\end{change2}

As final step, \tool constructs a \scade test case (Figure~\ref{fig:tool}, \emph{Test synthesizer}) for each of the selected C tests. It thus obtains a test suite in \scade format, which can be executed within the \scade test environment. 

Below we discuss in detail the design of the Driver synthesizer and the Test synthesizer.
\begin{change}
We then close this section by remarking the core original ideas that our test generator \tool settles in the analysis algorithm that it instructs with the Driver synthesizer.
The Test synthesizer is rather an engineering effort, though important to finalize the generated test suites. 
\end{change}

\subsection{The Driver Synthesizer}
\label{sec:driverSynthesizer}
The goal of the Driver synthesizer is to augment the C translation of the \scadelang model under test with an \emph{analysis driver}, designed to steer the symbolic analysis of the execution paths in the program. In pure technical terms, the analysis driver provides the entry-function that \klee shall symbolically execute in order to generate test inputs for the program under test. The analysis driver assigns 
\begin{change3}
symbolic values to the program inputs, 
\end{change3}
and then calls the target program one or multiple times with the symbolic inputs, to unfold the possible sequences of transitions of the state machine that the program represents.
Indeed each call to the target program corresponds to firing a transition of the state machine, and thus each execution path through the analysis driver corresponds to a sequence of state transitions that \tool \begin{change3}has to analyze symbolically.\end{change3}

The Driver synthesizer builds an analysis driver that steers \klee to symbolically execute (and thus generate test inputs for) the program paths and the execution sequences that satisfy the single-state-path-coverage (SSPC) testing criterion with respect to the state machine model implemented in the \scadelang program. The SSPC criterion requires to \begin{change3}
exercise
\end{change3}all paths and execution sequences that traverse states at most once~\cite{Pezze:SWTesting:book:2007}.   

The resulting analysis driver 
comes in the shape of a general algorithm, representing the overall steering strategy for satisfying the SSPC criterion, and a set of \begin{change2} automatically generated\end{change2} hook functions 
called from the general algorithm, representing the program-specific tailoring of the analysis driver. 
Algorithm~\ref{alg:driver} formalizes the general steering algorithm of the analysis driver in pseudo-code, with the calls of the hook functions represented within framed pseudo-code. 
The hook functions are either functions that already belong to the C program generated with \kcg, or functions that the Driver synthesizer 
generates and injects in the program.
The general algorithm  indicates that the analysis driver starts in a state that corresponds to the initial state of the program (line 6), with the program outputs 
\begin{change3}
initialized to default values by calling the initialization routine that \scadelang specifically generates as part of the code of each component
\end{change3}
(line 5), and considering an initially empty set of visited states (line 7). 
The hook function $default\_values$ (line 5) for initializing the program outputs with default values is part of the C program generated by \kcg.  
Then, the driver iterates through the loop at lines 8--14, where it calls the program under test once per iteration (line 10), until the execution of the program leads to an already visited state (line 8). Exiting the loop corresponds to an execution sequence that we must consider according the SSPC testing criterion and for which \klee will then generate a corresponding test input.

\begin{algorithm}[t]
\begin{algorithmic}[1]
\caption{The algorithm of the analysis driver}\label{alg:driver}

\LetBlock{
\\
$program$ be the program under test,\\
$inputs$ be a reference to the inputs of the program,\\
$outputs$ be a reference to the outputs of the program,\\
$s_0$ be the initial state of the program,
}
\Statex
\State $outputs \gets$ \fbox{$default\ values()$} \label{algo:init:outc}
\State $state \gets s_0$ 
\State $visited \gets \emptyset$   
\While{$state \notin visited$}
	\State $inputs \gets$ \fbox{$fresh\_symbols()$}
	\State $state', outputs \gets$ \fbox{$program(state, inputs, outputs)$}\label{algo:call-program} 
    \State \fbox{$save\_symbolic\_expressions(outputs)$}
    \State $visited \gets visited \cup \{state\}$
	\State $state \gets state'$
\EndWhile 

\end{algorithmic}
\end{algorithm}

At each iteration of the the loop at lines 8--14, the analysis driver triggers the possible state transitions of the \scadelang program by first assigning the program inputs with fresh symbolic values (line 9), and then symbolically executing the program under test to analyze the possible execution paths  (line 10). For each analyzed path, it saves the current symbolic values of the outputs to enable the \tool Test synthesizer to generate regression oracles later on (line 11), updates the set of visited states (line 12), and iterates with the analysis of the next state (line 13).
The hook functions at the first three steps inside the loop crucially depend on 
the restrictions that \begin{change3}\kcg\end{change3} enforces to foster dependable safety-critical software.
Below, we explain these hook functions in detail.  

\paragraph{Function fresh\_symbols (Algorithm~\ref{alg:driver}, line 9)} This hook function assigns the program inputs with symbolic values.
The Driver synthesizer generates the code of the hook function $fresh\_symbols$ based on the knowledge that \begin{change3}
\kcg does not generate
\end{change3}recursive data structures, dynamic memory allocation or \begin{change3}variable-length\end{change3} arrays. This guarantees the viability of unfolding all fields of primitive types that  belong to  the  input  data  structures at any nesting level, since these fields are necessarily a finite set. Thus, the analysis driver synthesizer customizes the code of function $fresh\_symbols$ such that it assigns each  primitive-typed input (received either as an input variable or as a field nested in an input data structure) to a fresh symbolic value, \begin{change2}while it initializes all pointer-typed inputs and arrays with references to concrete memory locations.\end{change2}\footnote{\begin{change2}
We remark that, by initializing  all pointers and arrays with references to concrete memory locations, our approach guarantees by-design that no memory access through a pointer can ever result into a \emph{symbolic memory access} (a memory access in which the memory location itself \begin{change3}-- or the base of the array --\end{change3} is a symbolic, non-deterministic value) during symbolic execution. Accessing arrays with symbolic indices can still lead to non-deterministic memory accesses, which \klee models with formulas expressed in the theory of arrays~\cite{GaneshD07}, \begin{change3}
consistently with the semantics of the program under test.
\end{change3} In this case, our approach guarantees that these formulas predicate on concrete, non-overlapping arrays of fixed size, which can be addressed without particular challenges with SMT solvers at the state of the art~\cite{GaneshD07}.\end{change2}} 
The Driver synthesizer uses the tool ANTLR4~\cite{bovet_parr_2008} to parse the type definitions in the C program for the sake of generating the code of function $fresh\_symbols$.

Let us consider, for instance, the working program that we introduced in Figure~\ref{fig:scade:program}. With reference to the corresponding C program of Figure~\ref{fig:c:program}, the Driver synthesizer
generates the hook function $fresh\_symbols$ as indicated in Figure~\ref{fig:fresh-symbols}.
By inspecting the considered C program, the analysis driver synthesizer
identifies that the data structure of type \texttt{inC\_WingMirrorControl\_CarControl}, which represents the program inputs, includes a field \texttt{ctrl} and a field \texttt{wingMirrorData}, respectively. The former field is defined as an enumeration type, i.e., a primitive type, and the latter field is an array, i.e., a non-primitive type.
Thus, the Driver synthesizer
inspects the definition of the array, revealing that it consists of two items of primitive type (again an enumeration).
The generated function $fresh\_symbols$ ultimately consists of C code that
initializes a new instance of the data structure in memory (Figure~\ref{fig:fresh-symbols}, line 2), relies on \klee (operation \textit{klee\_init}) to initialize the primitive  field \texttt{ctrl} with a new fresh symbol (line 3), initializes the non-primitive field \texttt{wingMirrorData} as a new array instance with two items (line 4), and initializes the two items in the array with further fresh symbols (lines 5 and 6).

The operation \textit{klee\_init} for initializing the inputs with fresh symbols takes two main parameters: one is the input to be initialized passed by reference, and the other one is a name (a string of characters) to be associated with that symbolic value. 
For instance, we might define the name "ctrl" for the fresh symbol that function $fresh\_symbols$ associates with the input \texttt{ret->ctrl} at line 2.  
Upon generating test inputs as possible concrete values of the symbols, \klee will use the provided name to indicate the input data to which those values refer.
We postpone to Section~\ref{sec:test-synthesizer} the discussion on how we specifically define the names for the fresh symbols to facilitate the task of synthesizing \scade test cases out of the test inputs obtained with \klee.

\begin{figure}[t]
\centering
\scriptsize
\begin{lstlisting}[language=C,numbers=left,stepnumber=1]
inC_WingMirrorControl_CarControl* fresh_symbols() {
    inC_WingMirrorControl_CarControl* ret = malloc(sizeof(...)); 
    klee_init(&ret->ctrl, "..."); 
    ret->wingMirrorData = malloc(2 * sizeof(...));
    klee_init(&ret->wingMirrorData[0], "...");
    klee_init(&ret->wingMirrorData[1], "...");
    return ret;
}
\end{lstlisting}
\caption{The hook function $fresh\_symbols$ for the sample program of Figure~\ref{fig:c:program}} 
\label{fig:fresh-symbols}
\end{figure}

\paragraph{Executing the program under test (Algorithm~\ref{alg:driver}, line 10)} The hook function $program$ at line 10 represents a call to the program under test, which is already part of the C code generated with \kcg.  The program receives the current state, the freshly initialized symbolic inputs and the current values of the outputs, and executes a state transition, possibly yielding a new state and new outputs. 
When executing $program$, \tool relies on the knowledge that the C translation of a \scadelang model consists (by construction) of all deterministic and terminating program paths, and thus the symbolic execution is guaranteed to terminate without need of enforcing any scope bound for the analysis.

\paragraph{Function save\_symbolic\_expressions (Algorithm~\ref{alg:driver}, line 11)} This hook function saves the symbolic expressions associated with the outputs after each execution of the program under test. This enables \tool to solve (at a later step) these expressions to concrete values \begin{change3} that consistently match with the selected inputs, and use those value to define regression oracles within the test cases\end{change3}. At the state of the art, generating regression oracles is a common functionality offered by most test generators~\cite{Fraser:EvoSuite:ESECFSE:2011}: a regression oracle defines the expectation that the outputs shall be equal to the values observed during the test generation process, which is trivially true when executing the test cases against the program that is being considered, but may provide meaningful insights on possible regressions against future new versions of the program.   

The same considerations that we discussed for function $fresh\_symbols$, related to the possibility of unfolding the primitive fields in the input data structures at any nesting level in finite steps, hold as well for function $save\_\-symbolic\_\-expressions$ with the only change that, in this case, the function unfolds the symbolic expressions associated with all primitive fields that belong to the output data structures of the program.  
For our working program, 
\begin{change3}
\tool
\end{change3}
customizes function $save\_\-sym\-bolic\_\-expressions$ to save the symbolic expressions associated with the primitive-typed output \texttt{carState}  and the two primitive outputs that comprise the array \texttt{mirrorCommand}.

Technically, function $save\_symbolic\_expressions$ generates a fresh symbol for each primitive-typed output, and informs \klee of the assumption that the new fresh symbol shall be equal to the value of the symbolic expression that is currently associated with the given output. This can be done with the  the \klee API \texttt{klee\_assume}. For instance, for saving the symbolic expression associated with the output \texttt{carState}, $save\_symbolic\_expressions$ generates a new fresh symbol (say $s$) and then calls \texttt{klee\_assume(s==carState)}. This leads \klee to compute a result for the symbol $s$ that reveals the value of \texttt{carState} at the moment when the assumption was evaluated during symbolic execution, consistently with the values that \klee computed for all other inputs.

\paragraph{Weak transitions}\label{sec:wtransitions}

We now discuss a refinement of the steering algorithm (Algorithm~\ref{alg:driver}) aimed to handle a special types of state transitions, called \emph{weak transitions}, which can be defined in \scadelang models. When a weak transition is fired, the actions that it defines are activated and the state is updated, but the outputs of the target state become active one execution cycle later.  Thus, a weak transition
requires two, rather than one, execution cycles to complete. During the second cycle the state machine stays in the destination state of the weak transition, that is therefore visited twice. 

Algorithm~\ref{alg:driver2} extends the analysis driver to handle weak transitions. This new algorithm is equal to Algorithm~\ref{alg:driver}, but includes the additional steps highlighted with gray-shadowed background. The algorithm has a new dependency on the predicate $weak\_transition(state_a, state_b)$ (line 5) that indicates whether or not the transition from $state_a$ to $state_b$ is a weak transition. We can deduce this information automatically out of the metadata that 
\scade associates with 
the program under test. 
After each execution step, if 
a weak transition is fired, i.e., if the predicate $weak\_transition(state, state')$ is true at line 17, then the variable $stutter$ memorizes the fact. In this case, at the next iteration, the destination state of the weak transition is not added to the set of the visited states (line 14). This allows the program to  complete the weak transition, which requires to  
visit that state once again, and then progress further on.

\begin{algorithm}[t]
\scriptsize
\begin{algorithmic}[1]
\caption{The algorithm of the analysis driver extended for weak transitions}\label{alg:driver2}

\LetBlock{
\\
$program$ be the program under test,\\
$inputs$ be a reference to the inputs of the program,\\
$outputs$ be a reference to the outputs of the program,\\
$s_0$ be the initial state of the program,\\
\colorbox{light-gray}{$weak\_transition$ be a predicate that is true for state-pairs that correspond to weak transitions.}
}
\Statex
\State $outputs \gets default\ values$ 
\State $state \gets s_0$ 
\State $visited \gets \emptyset$   
\State \colorbox{light-gray}{$stutter \gets false$}
\While{$state \notin visited$}
	\State $inputs \gets fresh\_symbols()$
	\State $state', outputs \gets program(state, inputs, outputs)$ 
    \State $save\_symbolic\_expressions(outputs)$
    \If{\colorbox{light-gray}{$\neg$ stutter}}
        \State $visited \gets visited \cup \{state\}$
    \EndIf
    \State {\colorbox{light-gray}{$stutter \gets$ weak\_transition(state, state')}}
	\State $state \gets state'$
\EndWhile 

\end{algorithmic}
\end{algorithm}

\subsection{The Test Synthesizer}\label{sec:test-synthesizer}

\begin{figure}[t]
\scriptsize
\begin{tabular}{l l c | c | c}
\multicolumn{3}{l|}{\textbf{Name of the fresh symbol (field, type, sequence)}} & \textbf{Test} & \textbf{enum}\\
\textbf{field} & \textbf{type} & \textbf{seq} & \textbf{input}& \textbf{value}\\
\hline\hline
inC.ctrl &   enum Lock &   1 &  0 & UNLOCKED\\
inC.wingMirrorData.automaticControl &   boolean &   1 &  false & - \\
inC.wingMirrorData.mirrorState[0] &   enum MirrorState &   1 &  0 & OPEN\\
inC.wingMirrorData.mirrorState[1] &   enum MirrorState &   1 &  0 & OPEN\\
outC.carState &   enum Lock &   1 &  0 & UNLOCKED\\
outC.mirrorCommand[0] &   enum MirrorState &   1 &  0 & OPEN\\
outC.mirrorCommand[1] &   enum MirrorState &   1 &  0 & OPEN\\
inC.ctrl &   enum Lock &   2 &  1 & LOCKED\\
inC.wingMirrorData.automaticControl &   boolean &   2 &  true & - \\
inC.wingMirrorData.mirrorState[0] &   enum MirrorState &   2 &  0 & OPEN\\
inC.wingMirrorData.mirrorState[1] &   enum MirrorState &   2 &  0 & OPEN\\
outC.carState &   enum Lock &   2 &  1 & LOCKED\\
outC.mirrorCommand[0] &   enum MirrorState &   2 &  1 & CLOSED\\
outC.mirrorCommand[1] &   enum MirrorState &   2 &  1 & CLOSED\\\hline
\end{tabular}\\
~\\
~\\
(a) The test inputs that \klee generated for an execution path (through the analysis driver)  for the sample \scadelang program of Figure~\ref{fig:scade:program}\\
\begin{lstlisting}[language=Java]
###################################################
## WingMirrorControl_WingMirrorFSM, Test case: 00002
####################################################

#Test step 1
SSM::set ctrl UNLOCKED
SSM::set wingMirrorData.automaticControl  false
SSM::set wingMirrorData.mirrorState {(OPEN,OPEN)}
SSM::check carState UNLOCKED
SSM::check mirrorCommand {(OPEN, OPEN)}
SSM::cycle 

#Test step 2
SSM::set ctrl LOCKED
SSM::set wingMirrorData.automaticControl  true
SSM::set wingMirrorData.mirrorState {(OPEN, OPEN)}
SSM::check carState LOCKED
SSM::check mirrorCommand {(CLOSED, CLOSED)}
SSM::cycle 
\end{lstlisting}
(b) The \scade test case synthesized out of the test inputs from \klee
\caption{A test case generated for the sample program of Figure~\ref{fig:scade:program}} 
\label{fig:testcase}
\end{figure}

\begin{change2}
The  \tool Test synthesizer uses the test inputs obtained with \klee to construct test cases in \scade format. Figure~\ref{fig:testcase}.b reports a sample test case in \scade format that was  generated with \tool. It consists of two test steps: The first test step sets (\textit{SSM::set} test statements) \texttt{ctrl} to \texttt{UNLOCKED}, \texttt{autom\-aticCon\-trol} to false and \texttt{mirrorState} to \texttt{OPEN} for both wing mirrors, in order to unlock the car and opening the wing mirrors. Thus, the test case doublechecks (\textit{SSM::check}) that, after this step, \texttt{carState} is equal to \texttt{UNLOCKED} and the outputs \texttt{mirrorCommand} are both assigned to \texttt{OPEN}.   
When the test case executes the statement  \textit{SSM::cycle}, \scade executes the test step and checks the values of the outputs accordingly. The second test step switches  \texttt{ctrl} to \texttt{LOCKED}, and \texttt{automaticControl} to true, then expecting that the  \texttt{carState} moves to \texttt{LOCKED} while issuing \texttt{mirrorCommand} outputs equal to \texttt{CLOSED}.

To synthesize the test cases in \scade format,  the \tool Test synthesizer  renders the test inputs that \klee yielded for a given execution path in the form of suitable \textit{SSM::set} test statements, and \begin{change3}renders the regression oracles that \klee yielded for that path in the form of suitable \textit{SSM::check} test statements.\end{change3} For the execution paths that \klee explored by issuing multiple calls of the program under test, the corresponding test cases shall include separate a test step (\textit{SSM::cycle}) for each program call, and
the Test synthesizer shall consistently map the test inputs that correspond to each program call with the
inputs of each step within the \scade test cases.
\end{change2} 

The Test synthesizer relies on a set of naming conventions that the analysis driver enforces when defining the names for the symbolic values. In detail, the analysis driver makes sure that the name of each fresh symbol specifies
\begin{inparaenum}[(i)]
\item the name of the input field initialized with the fresh symbol, 
\item the type of the input field, and
\item the sequence number of the program call for which the analysis driver instantiated the fresh symbol.
\end{inparaenum}

For instance, with reference to the the code of function $fresh\_symbols$ generated for our working program (Figure~\ref{fig:fresh-symbols}), the fresh symbol that the analysis driver associates with the input field \texttt{ctrl} at the second call of the program under test (for the execution paths that make at least two calls of the program) is named as
\textit{"field: inC.ctrl, type: enum Lock, sequence: 2"}. 
Thus, when \klee
yields a test input 1 for that symbol, the test synthesizer understands that the value 1 shall be assigned to the input field \texttt{ctrl} at the second program call made in the test case. Moreover, knowing that the field is of type \emph{enum Lock}, it can deduce that the value 1 refers to the second item defined in that enumeration, i.e., the value \texttt{LOCKED}. Thus,  the test synthesizer  generates the assignment \texttt{ctrl~=~LOCKED} at second test step.

Figure~\ref{fig:testcase} shows the test inputs (Figure~\ref{fig:testcase}.a) that \klee generates for an execution path through the analysis driver for  the sample \scadelang program of Figure~\ref{fig:scade:program}, and the \scade test case that \tool synthesizes correspondingly (Figure~\ref{fig:testcase}.b). 
The figure indicates the test inputs in tabular form to improve readability. Each row of the table corresponds to a test input from \klee. The first three columns represent the name that the analysis driver associated with the fresh symbol. As we described above, each symbol name is comprised of a field-, type- and sequence-specifier. The fourth column indicates the specific test input value that \klee yielded. The fifth column shows the matching enumeration value for test inputs of enumeration types.
The test inputs that correspond to the fields of the data structure \texttt{inC} were generated in the hook function $fresh\_symbols$ of the analysis driver: They indicate the input values for the test case. The ones that correspond to the fields of the data structure \texttt{outC} were generated in the hook function $save\_symbolic\_expressions$: They indicate values for regression oracles.

As the table indicates, \klee generated 14 inputs for the considered execution path. These 14 inputs refer to two subsequent calls of the program under test that occur within the execution path, as
the value of the sequence-specifier, either 1 or 2, indicates that the first 7 test inputs map to the first program call, and the following 7 test inputs map to the second program call, respectively.

Thus, \tool synthesizes a \scade test case \begin{change3}
consisting\end{change3} of two test steps (Figure~\ref{fig:testcase}.b). The first test step sets (\textit{SSM::set}) \texttt{ctrl} to \texttt{UNLOCKED}, \texttt{autom\-aticCon\-trol} to false and \texttt{mirrorState} to \texttt{OPEN}  for both wing mirrors. This results in unlocking the car and opening the wing mirrors, and in fact the test case defines \begin{change3}the regression oracles\end{change3} (\textit{SSM::check}) stating that this test step shall lead to a state in which the \texttt{carState} is equal to \texttt{UNLOCKED} and the outputs \texttt{mirrorCommand} are both set to \texttt{OPEN}.   
When the test case executes the statement  \textit{SSM::cycle}, \scade executes the test step and checks the values of the outputs accordingly. The second test step switches  \texttt{ctrl} to \texttt{LOCKED}, and \texttt{automaticControl} to true, then expecting in the assertions that the  \texttt{carState} moves to \texttt{LOCKED} while issuing \texttt{CLOSED} for both \texttt{mirrorCommand} outputs.

\subsection{Remarks}
The method that \tool realizes to initialize the program inputs with symbolic values, execute the program, and save the values of the outputs,
would hardly work if we were addressing the symbolic execution of an arbitrary C program. Thus, our design of the test generator \tool is tightly connected to the research hypotheses that this paper formulates about the class of programs identified by programming languages for safety-critical software, out of which we refer to \scadelang as a representative case. 

In detail, with reference to the hook functions $fresh\_symbols$ and $save\_sym\-bolic\_expressions$ that we introduced and discussed in this section, if the inputs and the outputs of the program could be defined of the type of dynamically allocated recursive data structures, the analysis driver that \tool synthesizes might lead \klee though infinite recursive steps in the attempt to initialize the fields at any nesting level, since the possible nesting levels would \begin{change3}be\end{change3} unbounded for a recursive data structure.  \begin{change3} If dynamic memory allocation had to be considered, symbolic execution should handle pointer-aliases for the pointers present in input data,\end{change3} by considering all the possible alternative initializations in which they could either hold null values, or refer to any compatible memory location that belongs to the input state~\cite{visser:generalizaed:2003}.
\begin{change3}If input arrays with non-statically-known length were allowed,\end{change3} there would be no immediate way to initialize them by unfolding their internal items.

With reference to the hook function $program$ (which executes the state transitions of the \scadelang program under test), \tool relies on the knowledge that the program under test is fully deterministic and does not include unbounded loops or recursion. This assumption guarantees that the analysis driver always analyzes a finite number of execution paths in each pass of the program, and always
terminates for each execution path, without need of specifying any customized bound neither in the target program, nor within the symbolic executor. 
In general, this is impossible for symbolic-execution-based test generators that address arbitrary programs.

 \section{Case Study}
\label{sec:caseStudy}
In this section we report on a case study where we evaluated the effectiveness of symbolic execution, 
as instantiated in our tools \tool described in Section~\ref{sec:generation},
for generating test cases for safety-critical software \begin{change3} developed in \scadelang.\end{change3}
 We considered a set of \scadelang programs developed as part of a project for an on-board signaling unit for high speed rail. This project is currently being developed by an industrial partner, \begin{change3}
with whom we are collaborating.\end{change3} We used \tool to automatically generate test cases for the considered \scadelang programs, and we evaluated our approach in terms of both the ability of \tool to successfully accomplish the test generation process, and the quality of the resulting test suites. 

Below we explain the research questions that drove our evaluation, describe the considered \scadelang programs, present the experimental setting of the case study,  report on the results, and discuss the main threats to the validity of our current conclusions. 

\subsection{Research Questions}
In the case study we aimed to answer the following research questions: 

\begin{itemize}
\item RQ1: Does \tool accomplish test generation within acceptable test budgets? 

\item RQ2: What is the quality of the test suites that \tool generates? 
\end{itemize}

We answer RQ1 by quantifying how many execution paths \tool actually analyzes when generating test cases for a set of \scadelang programs implemented by our industrial partner (presented below in Section~\ref{sec:subjects}), and how long it takes overall to complete the test generation process for those programs.
RQ1 aims to produce empirical evidence 
that we can effectively exploit symbolic execution to generate test cases for safety-critical software \begin{change3}in \scadelang.\end{change3} As we explained in Section~\ref{sec:generation}, \tool concretizes this hypothesis by tailoring its implementation of symbolic execution on the restrictions by which \begin{change3}\kcg\end{change3}fosters by-design safety guarantees in the programs. Thus, as RQ1 states, we aim to empirically confirm whether or not \tool, by its distinctive design, indeed succeeds to accomplish the test generation process within acceptable test budgets.

We answer RQ2 by evaluating the quality of the test suites that \tool generates for the considered programs. We evaluate the quality of the test suites (i)~in absolute terms, i.e., by measuring the size and the structural thoroughness of the test suites
\begin{change3}and by experiencing with the generated test suites to support component-level testing of the considered programs\end{change3},
(ii)~in comparison with the 
\begin{change}
manually derived test suites that were already available for three of the considered programs,
(iii)~in comparison with test suites automatically derived with a
search-based test generation approach~\cite{afl}.
\end{change}RQ2 aims to confirm the  merit of generating test cases based on symbolic execution.

\subsection{Subject Programs}
\label{sec:subjects}

\begin{table}[t]
\caption{Subject programs}
\label{tab:programs}
\begin{change}
\centering\scriptsize
\begin{tabular}{l|p{9cm}}
\textbf{Subject} & \textbf{\#Description}\\ 
 \hline\hline
shunting & Sorts  railway  vehicles into a complete train \\ \hline
dc\_1, dc\_2, ..., dc\_14 & Check data consistency of received messages\\ \hline 
radiohole & Deactivates radio connection supervision when train is in a  radio hole area \\ \hline
crossnonlx	& Monitors a level crossing area that is not protected by external authorities \\\hline
baliseinfo & Renders messages from on-railway transporders to the driver \\\hline
emergency\_1 &  Updates on-board data when receiving an emergency message	\\\hline
emergency\_2 & Acknowledges  radio control center when receiving an emergency message\\\hline
mema & Rejects movement authorities if there are emergency messages\\\hline
trackside & Receives and stores values from trackside equipments\\\hline
vbc & Updates the list of known transponders 
\\\hline
coordfromrbc &	Updates the coordinate system as specified by the ground control\\\hline
adfactordmi\_1 & Warns the driver if the railway adhesion factor is slippery \\\hline
adfactordmi\_2 & Renders the railway adhesion factor in the GUI	\\\hline
driveridins & Updates the driver ID as indicated through the GUI	\\\hline
eirene & Stores the EIRENE number as indicated through the GUI		\\\hline
ertmslevel & Updates the operating level  as indicated through the GUI		\\\hline
natvalues &	Verifies the national values of the currently traversed region 	\\\hline
networkidins &	Updates the identifier of the radio network \\\hline
rbcidins	& Stores the ID of the radio control center ID as indicated through the GUI \\\hline
trainDataUpdate	& Updates the train data stored on board\\\hline
trainDataInsertion & Inserts new train data among the ones stored on board	\\\hline
message129 & Notifies changes of train data to the radio control center
\\\hline
runnumber\_1 & Updates the train ID on board\\\hline
runnumber\_2 & Notifies changes of the train ID to the radio control center
\\\hline
\hline
\end{tabular}
\end{change}
\end{table}

\begin{table}[t]
\caption{Statistics of the subject programs}
\label{tab:metrics}
\begin{change}

\centering\scriptsize
\begin{tabular}{l||r|rr|r|r||r|}
 & \multicolumn{5}{c||}{\textbf{\scadelang model}} & \multicolumn{1}{c|}{\textbf{C code}} \\
\textbf{Subject} & \textbf{\#States} & \multicolumn{2}{c|}{\textbf{\#Transitions}} & \textbf{\#Inputs} & \textbf{\#Outputs} & \textbf{LOC$^{(*)}$} \\ 

 & & \textbf{weak} & \textbf{strong}&  &  & \\ 
 \hline
shunting & 5 & 2 & 8 & 12 & 14 & 646 \\ 
dc\_1 & 1 & 1 & - & 13 & 7 & 175 \\ 
dc\_2 & 1 & 1 & - & 1 & 2 & 43 \\ 
dc\_3 & 1 & 1 & - & 5 & 3 & 95 \\ 
dc\_4 & 1 & 1 & - & 3 & 4 & 62  \\ 
dc\_5 & 1 & - & 1 & 3 & 1 & 32 \\ 
dc\_6 & 1 & 1 & - & 3 & 4 & 67  \\ 
dc\_7 & 1 & - & 1 & 3 & 1 & 32 \\ 
dc\_8 & 1 & - & 1 & 2 & 1 & 30 \\ 
dc\_9 & 1 & 1 & - & 5 & 15 & 464 \\ 
dc\_10 & 1 & 1 & - & 3 & 9 & 239 \\ 
dc\_11 & 1 & 1 & - & 1 & 3 & 69 \\ 
dc\_12 & 1 & 1 & - & 14 & 17 & 96 \\ 
dc\_13 & 1 & 1 & - & 3 & 7 & 67 \\ 
dc\_14 & 1 & - & 1 & 1 & 1 & 35 \\ 
radiohole & 3 & 2 & 1 & 2 & 2 & 361 \\ 
crossnonlx	&3	&2	&1&	6&	4&	556 \\
baliseinfo&	1&	1&	0&	1&	2&	147\\
emergency\_1&	1&	1&	0&	9&	4&	865\\
emergency\_2&	1&	1&	0&	9&	6&	711\\
mema&	1&	1&	0&	4&	1&	798\\
trackside&	1&	1&	0&	3&	0&	225\\
vbc&	1&	1&	0&	7&	1&	1,011\\
coordfromrbc&	1&	1&	0&	1&	1&	366\\
adfactordmi\_1&	1&	1&	0&	3&	1&	125\\
adfactordmi\_2&	1&	0&	1&	1&	1&	54\\
driveridins&	1&	1&	0&	1&	1&	262\\
eirene&	1&	0&	1&	3&	1&	124\\
ertmslevel&	1&	0&	1&	1&	1&	109\\
natvalues&	1&	0&	1&	1&	1&	265\\
networkidins&	1&	0&	1&	1&	1&	109\\
rbcidins	&1	&1	&0&	1&	1&	189\\
trainDataUpdate	& 1	&1	&0&	2&	19	&136\\
trainDataInsertion &	1&	0&	1	&2&	1&	291\\
message129&	1	&1&	0&	5	&1	&353\\
runnumber\_1&	1&	1	&0&	1&	1	&154\\
runnumber\_2&	1&	1&	0&	4&	1&	116\\
\hline

\end{tabular}

~\\$^{(*)}$ C code LOC values refer to the lines of code in the C functions specific of each \scadelang program, but each program includes more than 8,000 additional lines of code of data-type declarations, \begin{change3}
which define the data structures that comprise the inputs and the outputs of the programs.\end{change3} \end{change}
\end{table}

We considered the 
\begin{change}
37 \scadelang programs 
 listed in  Table~\ref{tab:programs}. The table defines an identifier (first column) that we use to refer to each subject program in the sequel of the paper, and provides a short description (second column) of the task that each program executes.  
 \end{change} 
These programs are part of the on-board signaling unit for high speed rail that our industrial partner is currently developing. For example, the first program, \texttt{shunting} implements the Shunting procedure. 
In the railway terminology, shunting is the process of sorting railway vehicles into complete trains. When a train is in \emph{shunting mode}, the on-board unit is responsible for the supervision of the speed limit that is allowed during the shunting operations, and to stop the train when it passes the defined border of the shunting area.
The shunting procedure that we consider as subject program shall handle the messages that the train receives from both the driver and the ground signaling equipment, to make decisions on when activating or deactivating the shunting mode. 
\begin{change}
The other 
programs
implement several control tasks, as checking and verify the consistency of the data that the on-board unit receives from the ground components,
computations of information for monitoring and controlling the train, 
rendering appropriate messages to the driver, and sending commands to the actuators.

\end{change}

Table~\ref{tab:metrics}  summarizes the main statistics on the internal structure of the subject programs, i.e.,  
the number of the states (column \textit{\#States}) and state transitions (columns \textit{\#Transitions}) of the state machine that corresponds to each \scadelang program, the number of inputs (column \textit{\#Inputs}) and outputs (column \textit{\#Outputs}) of each \scadelang program, and the number of lines of C code that correspond to each program after exporting it with \kcg. 
For the state transitions, the table reports separately the number of weak and strong (non-weak) transitions,  since the weak transitions count double in the sequences of transitions that \tool analyzes, as 
we explained in Section~\ref{sec:wtransitions} (Algorithm~\ref{alg:driver2}). The lines of C code refer to the code within the C functions that specifically correspond to each \scadelang program, without counting the lines of code of the data-type definitions in those programs. In fact, each program includes more than 8,000 further lines of code that define the data-types used in the C functions, and which \tool parses with ANTLR4 to instantiate the hook functions of the analysis driver.

For instance, the \scadelang implementation of \texttt{shunting} is a state machine with 5 states, 2 weak transitions and 8 strong transitions,
in which the states and the transitions 
are based on computations and conditions that involve 12 input and 14 output variables, respectivey, including the variables that represent the messages received and sent from on-board unit.
\begin{change}
Many subjects (all but \texttt{shunting}, \texttt{radiohole} and \texttt{crossnonlx})
\end{change}
implement computations that the on-board unit shall keep on repeating at each execution cycle, and thus they consist of a single state transition which represents the execution of the computation, and which keeps the program always in the same state. 
For instance, 
the \texttt{dc\_1..14} programs implement
data consistency checks that the on-board unit shall perform at each execution cycle.
These programs define either a weak or a strong transition according to whether or not, respectively, the check that they implement depends on feedback loops with their own outputs.

At the level of the C code, 
the considered programs range 
\begin{change}
between 30 and 
1,011 lines of code 
\end{change}
(plus the code defining the data types, i.e., as said, more than 8,000 additional lines of code) \begin{change}
being program \texttt{dc\_8} and 
program \texttt{vbc} the smallest and the largest program,
\end{change}
respectively.

\subsection{Experimental setting}
\label{sec:setting}
Our case study consisted of a set of experiments, one for each of the subject programs listed in Table~\ref{tab:metrics},  in which we ran \tool to generate test cases for the subject programs, executed the test cases in the \scade Suite and collected model coverage data.

We ran \tool on cloud facility hosted at our university, using a virtual machine equipped with Linux Ubuntu, 48 cpus, 150 GB of ram memory, which allowed for running multiple instances of \tool in parallel.

We handled the \scadelang programs with \scade Suite Version 2020 R2, which includes the corresponding version of the \kcg compiler that we use to obtain the C version of the subjects programs. We executed the test cases with the tool \scade Test Version 2020 R2.

During the experiments, for each subject program, we tracked
the number of paths that \tool identified during the symbolic execution phase, 
measured the time that it took to complete the test generation process, counted the number of test cases that it generated, and computed the model coverage that the test cases achieve against the \scadelang programs.

For measuring the model coverage of the test cases we relied on the  \scade Test tool, which automatically computes the model coverage while executing the test cases.
The coverage computed with \scade Test refers cumulatively to the portion of
executed states, and the modified condition/decision coverage of the transition guards.

In the case of programs \texttt{shunting}, 
\begin{change2}
\texttt{radiohole} and \texttt{crossnonlx} we were able to compare the test cases generated with \tool with  manually selected test suites that were already available for those programs at the time of our experiment. 
We \end{change2} compared the manual and the automatic test suites with respect to their difference in model coverage, focusing on the items that either test suite covers and the other one does not.  

\begin{change2}
For all other subject programs, the engineers at our industrial partners decided to rely directly on our tool (as \tool in fact became available while those programs were being implemented), aiming to optimize their effort for designing and implementing the test cases for those programs. 
\begin{change3}
To this end, they augmented the test cases generated with \tool with (manually defined) assertion-style test oracles, aiming to obtain test suites that could be readily used for component-level testing of the considered programs (other than for future regression testing of those programs).
This resulted in a semi-automatic approach to component-level testing empowered by our tool \tool, and allowed us to further validate the quality of the test suites generated with \tool in terms of usefulness for detecting component-level failures in the context of our industrial project.
\end{change3}

We remark that, on one hand, this choice of our partner affected  
our ability to extensively crosscheck the differences in effectiveness of automatically and manually generated test suites, respectively, since no manual test suite existed to compare with for any subject program but \texttt{shunting}, \texttt{radiohole} and \texttt{crossnonlx}; On the other hand we believe that the choice of dismissing fully manual testing  in favour of \begin{change3} working with
semi-automatic test cases (obtained by enriching with assertions the ones generated with \tool) \end{change3} supports the positive perception of our industrial partner on the effectiveness of our approach.
\end{change2}

\subsection{Results}
\label{sec:results}

\begin{table}[t]

\centering\scriptsize
\caption{Results of \tool for the subject programs considered in our case study} 
\label{tab:results}

\begin{change}
\begin{tabular}{l|| r|r||r|r| r|r}
 & & & & \multicolumn{2}{c|}{\textbf{\#test steps}} \\
\textbf{subject} & \textbf{time (s)} & \textbf{\#paths}  & \textbf{\#tests} & \textbf{avg} & \textbf{max} & \textbf{coverage}\\ \hline
shunting                & 286 & 3,367 & 20 & 3 & 5 & 86\% \\ 
dc\_1                   & 2 & 616  & 8 & 2 & 2 & 91\% \\ 
dc\_2                   & $<$1 & 2   & 2 & 2 & 2 & 100\%  \\ 
dc\_3                   & $<$1 & 16  & 6 & 2 & 2 & 100\%  \\ 
dc\_4                   & $<$1 & 3  & 2 & 2 & 2 & 92\% \\ 
dc\_5                   & $<$1 & 4 & 2 & 1 & 2 & 89\%  \\ 
dc\_6                   & $<$1 & 3 & 2 & 2 & 2 & 90\% \\ 
dc\_7                   & $<$1 & 4 & 2 & 1 & 2 & 80\% \\ 
dc\_8                   & $<$1 & 4  & 3 & 1 & 2 & 83\% \\ 
dc\_9                   & 2 & 208 & 9 & 2 & 2 & 100\% \\ 
dc\_10                  & $<$1 & 64 & 9 & 2 & 2 & 93\%\\ 
dc\_11                  & $<$1 & 3 & 2 & 2 & 2 & 100\% \\ 
dc\_12                  & $<$1 & 3  & 3 & 2 & 2 & 72\%\\ 
dc\_13                  & $<$1 & 20  & 4 & 2 & 2 & 98\% \\ 
dc\_14                  & $<$1 & 4  & 2 & 1 & 1 & 82\% \\ 
radiohole               &	117 &	45 &	6 &3	 & 3 &  95\%  \\
crossnonlx              &647	&294&	13   &  3& 3 &	84\%\\
baliseinfo              &1&	3&	3& 1& 2 & 	97\%\\
emergency\_1            &15&	28&	14&1 & 2 &	94\%\\
emergency\_2            &29	&8&	6&	1 & 2 & 82\% \\
mema                    &23&	17&	7& 1& 2 &	89\% \\
trackside               &1137&	3&	3&1 &2 & 	99\%\\
vbc                     &164	&77&	12&1 & 2 & 	94\%\\
coordfromrbc            &41	&7	&5&	1 & 2 & 83\% \\
adfactordmi\_1          &1860&	3&	3& 1& 2 & 	85\% \\
adfactordmi\_2          &1&	2&	2& 1& 1 & 	96\% \\
driveridins             &5&	10&	10&	1 & 2 & 89\% \\
eirene	                &3&	3	&3 & 1 & 1	&94\% \\
ertmslevel	            &2&	3&	3& 1& 1 & 	94\%\\
natvalues	            &1230&	4&	4&1 &  1& 	90\% \\
networkidins	        &1	&3	&3 & 1 & 1	&94\% \\
rbcidins                &3&	4&	4&	1 & 2 & 95\%\\
trainDataUpdate         & 47&	2&	1& 1& 2 & 	89\% \\
trainDataInsertion      &28	&4	&3&1 & 1 & 	95\%\\
message129          	&99&	80	&10	 & 1 & 2&83\% \\
runnumber\_1            &2&	3&	3& 1& 2& 	94\% \\
runnumber\_2            &3&	22&	7& 1& 2 & 	92\%\\
\hline
\end{tabular}
\end{change}
\end{table}

Table \ref{tab:results} summarizes the data on the execution of \tool in our experiments, and the test cases that it generated. 
For each subject program (column \textit{program}), the table reports the time in seconds taken to complete the overall test generation process (column \textit{time}), the number of execution paths analyzed with symbolic execution (column \textit{\#paths}), 
the number of test cases generated after executing the minimization step (column \textit{\#tests}), the average and maximum number of test steps within the test cases (columns \textit{\#test steps}),  and the model coverage of the test cases (column \textit{coverage}).

\subsection*{Test budget requirements (RQ1)}
Overall, the data in Table \ref{tab:results} support a positive answer to the research question RQ1 on whether \tool accomplishes the test generation process within acceptable test budgets. 
Furthermore, as \tool completed in finite time in all experiments, these data also 
support our hypothesis 
that, thanks to the language restrictions that \scade embraces to promote safe programs, we can exploit symbolic execution to efficiently explore the execution space of the programs under test 
without need of specifying custom bounds for the analysis. 

In detail, for most subject programs,
\tool took a few seconds to complete the test generation process. It took more than 1 minute only for 8 out of 37 subject programs, and more than 10 minutes only for 4 programs, namely, \begin{change3}
\texttt{crossnolx}, \texttt{trackside}, \texttt{adfactordmi\_1} and  \texttt{natvalues}, the maximum time being 31 minutes (1,860 seconds) in the experiment with program \texttt{adfactordmi\_1}.\end{change3}

In all experiments \tool used most computation time to complete the symbolic execution with \klee, under the guidance of the \tool analysis driver, while the other phases of \tool, i.e., synthesizing the analysis driver, and synthesizing the test cases in \scade format, took negligible time.

\begin{change2}
We investigated in further detail the experiments in which 
the time budget was not justified by the (low) number of symbolically executed paths. For these cases, we investigated whether the time budget was bounded by some complex execution conditions that took long time for the constraint solver to compute the solutions. To this end, we logged the number of queries that the symbolic executor issued to the constraint solver, and the queries for which the constraint solver took more than a specified time.
Table \ref{tab:SymbolicValues} shows these data in particular for the subject programs (column \emph{subject}) for which \tool executed for a number of seconds (column \emph{time}) higher than the number of symbolically analyzed execution paths (column \emph{\#paths}). The table reports the number of the queries issued in total to the solver (column \#queries) and restricted to the ones that took more than a millisecond to be solved (column $>$1).
As the table shows, indeed no query took more than a millisecond, confirming that the execution conditions generated during the analysis of the \scade programs result in simple constraint solving problems.
For these programs we were able to map the execution time to the large data structures that comprise their inputs, which required the initialization and the handling of many symbolic values during symbolic execution. 
\end{change2}
\begin{table}[]
\centering \scriptsize
\begin{change}

\caption{Data on the queries issued to the constraint solver}
\label{tab:SymbolicValues}

\begin{tabular}{l|rrrc|}
\textbf{subject}   & \textbf{time (s)} & \textbf{\#paths} & \textbf{\#queries} & \textbf{\textgreater{}1 ms} \\ \hline
radiohole          &	117 &	45 & 484 & 0  \\
crossnonlx         &    647	& 294  & 502	 & 0\\
emergency\_2       & 29      & 8     & 280   & 0      \\
mema                &23 & 17 &  122& 0 \\
trackside          & 1137    & 3     & 1305  & 0        \\
vbc                & 164     & 77    & 279   & 0        \\
coordfromrbc        &41	     & 7	& 155    &0 \\
adfactordmi\_1     & 1860    & 3    & 2530  & 0           \\
natvalues          & 1230    & 4     & 1296  & 0        \\
trainDataUpdate    & 47      & 2     & 215   & 0       \\
trainDataInsertion & 28      & 4     & 121   & 0       \\
message129         & 99      & 80    & 133   & 0       \\ \hline
\end{tabular}
\end{change}
\end{table}

\subsection*{Quality of the Test Suites (RQ2)}

Table \ref{tab:results} shows that the test suites that \tool generated in our experiments consist of a minimum of 1 test case, for program \texttt{trainDataUpdate}, up to a maximum of 20 test cases for \texttt{shunting}. The number of corresponding test steps is either 1 or 2 in all test cases generated for the programs that consist of only a strong or a weak transition, respectively, 
\begin{change2}
while it is higher for the three programs that define \scade models with more states and transitions, i.e.,  \texttt{shunting} (5 states, 10 transitions), \texttt{radihole} (3 states, 3 transitions) and \texttt{crossnonlx} (3 states, 3 transitions). For these programs, the generated test cases consist of  3 test steps on the average, up to a maximum of 5, 3 and 3 test steps for \texttt{shunting}, \texttt{radihole} and \texttt{crossnonlx}, respectively.   
\end{change2}

\begin{change2}
The generated test suites achieved a model coverage of 100\% for 4 subject programs, at least 90\% for 19 further programs, at least 80\% for 13 programs, 
\end{change2}
and 72\% in the only case of program \texttt{dc\_12}.

\paragraph*{Uncovered Items}
We inspected the programs with uncovered items in further detail, to investigate the reason why \tool missed the generation of test cases that cover those items. We tracked the uncovered items to four distinct motivations:

\begin{itemize}
\item 
\begin{change2}
Items that depend on infeasible program paths: In fact, many subject programs include infeasible paths, the most frequent case being the one of programs structured with some (sub-)procedures, where the procedures define general algorithms, but the program calls them only in  specialized contexts (e.g, with constant values passed for some parameters)
and thus inhibits the possibility of executing some branches (e.g., the branches that depend on parameter values different than the used constants). 
\end{change2}

\item 
\begin{change2}
Unreported coverage: The \scade Test tool does not report the coverage of the items that, although executed during the test cases, do not map to any observable output of the \scade operators in the programs under test.
In the considered programs, this happens for a set of operators defined to update stored data: these operators take an input, and use it to do the update, without producing any explicit output. 
This leads to the \scade test tool to misleadingly classify some items of our subject programs as uncovered. As we are discussing with our partner, this observation calls for some refactoring of the mentioned operators, to improve the precision of the coverage measurements.

\end{change2}

\item Functional behaviors out of the scope the single-state-path-coverage testing criterion that \tool uses for steering the test generation process:
We observed  uncovered functional behaviors in program \texttt{shunting}. 
The \scade model of this program includes two 
model states in which the train expects a message from the ground equipment. These states  implement the \emph{degraded behavior} of assuming that the ground equipment is not responding, if the expected message is not received within a specific number of execution cycles. 
As a matter of facts, these behaviors correspond to execution sequences that iterate in the same state for multiple execution cycles, and are thus out of the scope of the single-state-path-coverage testing criterion that \tool is designed to satisfy.

\item Uncovered modified condition/decision targets: As we commented in Section~\ref{sec:generation} while discussing the Miminizer step of \tool, a limitation of the current implementation is to select test cases based on statement coverage, which is a grosser grained criterion than the modified condition/decision coverage of transition guards considered in \scade Test. This resulted in a few uncovered modified condition/decision targets in the current experiments, even if \tool analyzed all execution paths. As said, we aim to overcome this limitation of \tool in future release.

\end{itemize}

\begin{change2}
Out of the above cases, only the last two map to limitations of our approach.  While  the former of these limitations  suggests the strategy of complementing the automatically generated test cases for the programs with missing coverage (by searching for functional behaviors that require iterating multiple times though the same state), the latter could be mitigated by improving the implementation of \tool.
\end{change2}
\begin{change3}
We evaluated the room for the coverage improvement that we might achieve with a different strategy for selecting test cases out of the symbolically analyzed execution paths. To this end, we re-executed \tool after disabling the  Miminizer option in \klee, thus making \tool compute exactly one test case for each symbolically analyzed execution path. Table~\ref{tab:minimizer} compares the number of test cases and the coverage results that we achieved with and without the Minimizer, respectively (but for program \texttt{shunting} for which the high number of test cases computed without Minimizer -- 3,367 test cases -- exceeded the capability of \scade Test to execute the test suite). In the table, we highlighted in bold the 6 cases in which the coverage rate improved without using the Minimizer. The amount of improvement was 1\% for \texttt{radiohole} and \texttt{message129}, 2\% for \texttt{crossnolx} and \texttt{emergency\_2}, 4\% for \texttt{mema}, and up to 7\% for \texttt{dc\_1}. 
\end{change3}

\begin{table}[t!]

\centering\scriptsize
\begin{change3}
\caption{Results of \tool with and without the Minimizer} 
\label{tab:minimizer}
~\\
\begin{tabular}{l||rr|| rr|}
 & \multicolumn{2}{c||}{\textbf{with Minimizer}} & \multicolumn{2}{c|}{\textbf{no Minimizer}} \\
\textbf{subject} & \textbf{\#tests} & \textbf{coverage} & \textbf{\#tests} & \textbf{coverage} \\ \hline
shunting             & 20 &  86\% & 3,367&n.a. \\ 
dc\_1                  & 8 & 91\% &616  &\bf 98\% \\ 
dc\_2                     & 2  & 100\%&2  &100\%  \\ 
dc\_3                   & 6 & 100\% & 16 &100\% \\ 
dc\_4                   & 2 & 92\%&3 &92\%  \\ 
dc\_5                   & 2 & 89\%&  4& 89\%  \\ 
dc\_6                   & 2 & 90\%&  3& 90\% \\ 
dc\_7                  & 2 & 80\%&  4&80\%  \\ 
dc\_8                  & 3 & 83\%&  4&83\%  \\ 
dc\_9                    & 9 & 100\%&  208&100\%  \\ 
dc\_10                  & 9 & 93\%&  64&93\% \\ 
dc\_11                  & 2 & 100\% &  3&100\% \\ 
dc\_12                   & 3 & 72\% &  3&72\% \\ 
dc\_13                   & 4 & 98\% &  20&98\% \\ 
dc\_14                  & 2 & 82\% & 4 &82\% \\ 
radiohole              &	6 &  95\% &45  &\bf 96\% \\
crossnonlx             &	13  &	84\%& 294 &\bf 86\% \\
baliseinfo             &	3& 	97\%&  3&97\% \\
emergency\_1            &	14 &	94\%& 28 &94\% \\
emergency\_2            &	6 & 82\%& 8 &\bf 84\% \\
mema                    &	7 &	89\%& 17 &\bf 93\% \\
trackside               &	3 & 	99\%& 3 & 99\%\\
vbc                     &	12 & 	94\%&  77&94\% \\
coordfromrbc            &5& 83\%& 7 &83\% \\
adfactordmi\_1         &	3 & 	85\%& 3 &85\% \\
adfactordmi\_2         &	2& 	96\%&  2&96\% \\
driveridins             &	10 & 89\%&  10&89\% \\
eirene	                	&3 	&94\%&  3&94\% \\
ertmslevel	            &	3& 	94\%&  3&94\% \\
natvalues	            &	4& 	90\%&4  &90\% \\
networkidins	        &3 &94\%&  3&94\% \\
rbcidins               & 4& 95\%&  4&95\% \\
trainDataUpdate         &	1 & 	89\%& 2 &89\% \\
trainDataInsertion      &3 & 	95\%&  4& 95\%\\
message129          	&10	 &83\%& 80 &\bf 84\% \\
runnumber\_1            &	3& 	94\%&  3&94\% \\
runnumber\_2            &	7& 	92\%&  22& 92\%\\
\hline
\end{tabular}

~\\cov = n.a., if \scade Test failed due to too many test cases
\end{change3}
\end{table}

\paragraph*{Comparison with Manually Derived Test Cases}

In the case of the subject programs \texttt{shunting},
\begin{change2}
\texttt{radiohole} and \texttt{crossnonlx} we were able to compare the test cases generated with \tool with  manually selected test suites 
\end{change2}
that were already available for those programs at the time of our experiment. 
These test suites were designed  
in a functional fashion based on the  software requirements specified for the program, using the model-based test criterion of executing at least once all non-cyclic paths of the state machine and all conditions involved in the state transitions.   
\begin{change2}
The engineers reported to us that the analysis of the requirements, the selection of the test cases and their manual implementation in a the test suite took overall \begin{change2} 16 man-hours (two days of work), 3 man-hours (about half day) and 9 man-hours (about one day)  for \texttt{shunting}, \texttt{radiohole} and \texttt{crossnlnlx} respectively. \end{change2}They tracked the main challenges to i)~devising a suitable functional partitioning of the relevant cases to be tested (which in turn required to reiterate multiple times the study and the analysis of the requirement documents), ii)~analyzing the implementation to identify suitable input and test step sequences for exercising the identified set of relevant cases, and iii)~rendering the test cases in the specific language and format required by the \scade test tool (that we exemplified in Figure~\ref{fig:testcase}.b).

\begin{table}[th]
\centering\scriptsize
\caption{Comparison between automatically and manually derived test suites}
\label{tab:CoverageComparison}
\begin{change2}
\begin{tabular}{l||rrr||rrr|}
 & \multicolumn{3}{c||}{\textbf{Manual test suite}} & \multicolumn{3}{c|}{\textbf{TECS}} \\ 
\textbf{subject} & \textbf{time} & \textbf{\#tests} & \textbf{coverage} & \textbf{time} & \textbf{\#tests} & \textbf{coverage} \\ \hline
shunting   & 16 h & 15  & 95\% & 286 s & 20  & 86\%\\ 
radiohole  & 6 h  & 1   & 94\% & 117 s  & 6   & 95\%\\ 
crossnonlx & 9 h  & 3   & 80\%   & 647 s & 13  & 84\% \\ \hline
\end{tabular}
\end{change2}
\end{table}

Table \ref{tab:CoverageComparison} reports the main statistics of the manual test suites (columns \emph{Manual test suite}) for the three considered programs, sided to the statistics of the test suites that \tool generated (columns \emph{\tool}) for each of the programs. For each test suite we report the time taken to generate the test suite (column  \emph{time}), the number of test cases  (column \emph{\#tests}) and the corresponding model coverage (column \emph{coverage}).

The manually derived test suites are sightly more compact in terms of number of test cases than the automatically generated counterparts, but it is clear that pay higher costs in terms of working effort (several hours) in comparison with the relatively shorter time that developers must wait to obtain the test cases with \tool. 
In terms of coverage, the manual test suite of \texttt{shunting} achieves higher model coverage than the test suite that \tool generated for this program, but \tool achieved higher model coverage than 
the manual test suites for \texttt{radiohole} and \texttt{crossnonlx}.
\end{change2}

We analyzed the difference in the coverage data, focusing in particular on the items of the coverage domain that either test suite hits and the other one does not. 
In detail, for \texttt{shunting}, the manually designed test suite successfully executed the degraded behaviors (since they correspond to a specific transitions indicated in the requirements) that \tool missed 
as we already commented above.
On the other hand, the manually designed test suite missed some possible combinations of the conditions that participate in the transition guards, some of which were hit with \tool thanks to the systematic analysis of all execution paths in the program.
\begin{change}
Instead, we did not find any manually tested  behavior that \tool did not cover in \texttt{radiohole} and \texttt{crossnonlx}, where \tool was in fact able to cover some additional rare combinations.

\end{change}

\medskip
In summary, our case study indicated that the test generation approach that we propose in this paper, as instantiated in the tool \tool, successfully exploits symbolic execution to generate high-quality test suites for safety-critical programs in \scade, thus confirming the main research hypothesis of this paper. The test suites that we automatically generated with \tool in our experiments readily satisfied most domain-relevant test objectives, unveiling at the same time small portions of test objectives that require dedicated handling.
This straightforwardly suggests a combined approach in which the testers of safety-critical software can efficiently start working with the  test cases automatically computed with \tool, and then complementarily concentrate on the yet-missed behaviors, thus crucially improving both costs and the effectiveness of their test-design efforts.

\paragraph*{\begin{change3}Usefulness of \tool for semi-automatic component-level testing\end{change3}}
\begin{change3}
To further investigate the quality of the test suites generated with \tool, we worked jointly with our industrial partner to exploit those test suites for component-level testing of the considered programs. To this end, the test suites generated with \tool were augmented with assertion-style test oracles defined by test engineers based on the documented requirements, thus resulting in a semi-automatic approach to generating the component-level test cases. 
Manually adding the assertions took limited effort, a few minutes per test case: It required the test engineers to crosscheck the concrete inputs already provided in the test cases with the expectations defined in the requirement documents. This, we remark, is a radically simpler task than the manual effort quantified in Table~\ref{tab:CoverageComparison}, to design and implement the test cases from scratch, which encompasses a very much larger set of time consuming activities (such as, identifying a functional partitioning out of the requirements, devising suitable test steps and inputs, and implementing the \scade test cases from scratch). 
   
Table \ref{tab:faults} describes the faults that we identified by executing the test suites obtained in this way. Overall we revealed 7 previously unknown faults in four of the subject programs considered in our experiment. 
These results support the usefulness of the test suites generated with \tool for component-level testing.
\end{change3}

\begin{table}
\begin{change3}
\caption{Faults identified in the subject programs considered in our case study}
\label{tab:faults}
\scriptsize

\begin{tabular}{lp{8.4cm}}

\textbf{subject} & 
\textbf{fault}\\ 
 \hline
dc\_10  
& Wrong amount of data written in a queue\\

&  Wrongly defined algorithm\\ 

coordfromrbc 
& Missing update of a state variable\\ 

& Array updated with index starting at second (instead of first) item\\

emergency\_1 
& Output value out of expected range\\

&  Wrongly defined algorithm\\

emergency\_2 
&  Interrelated variables updated in wrong sequence
\\\hline
total & 7 faults\\
\end{tabular}
\end{change3}

\end{table}

\paragraph*{Comparison with search-based testing}

\begin{change2}
We investigated if our approach could work also by using search-based random testing heuristics in place of symbolic execution. To this end, we implemented an alternative version of \tool that used the test generator AFL~\cite{afl} instead of \klee to produce the test inputs. AFL is a test generator that is very popular for security vulnerability testing: it starts by performing random mutations on a set of (seed)  inputs provided by developers, and then progresses in search-based fashion by considering the newly generated inputs that increase code coverage as additional seeds. In our setting we executed AFL on the analysis-driver programs generated by \tool, providing initial seeds that included an input value for each program input that \tool handled symbolically when using \klee: \begin{change3}For each subject program, we seeded AFL with the input values from the the first test case that we had generated when using \klee.\end{change3}

The task of AFL was then to discover (by means of its search-based heuristics) further input values, as needed to cover the branches of the program under test.
\begin{change3}
Technically, we exploited the feature of AFL to feed back its own test generation mechanism with the test cases that execute new branches. Upon identifying test inputs that make the program execute new branches, AFL saves those test cases in a queue, aiming to consider them as possible seeds at next steps. Thus, 
for each subject program, we proceeded as follows: we executed AFL for 5 hours; We used our tool to translate the test cases in the final queue into test cases in \scade format; We executed the test cases with \scade Test to collect the corresponding coverage data. 
\end{change3}
We also repeated each test generation attempt 3 times to control for the random characteristics of AFL.

\begin{table}[t!]
\centering
\scriptsize
\caption{Comparison between \tool and AFL}
\label{tab:AFL}
\begin{change3}
\begin{tabular}{l||rr||rr||rr|}
& \multicolumn{2}{c||}{\textbf{TECS}} & \multicolumn{2}{c||}{\textbf{AFL}}&  \\  
\bf subject   & \bf \#tests         & \bf coverage         & \bf \#tests         & \bf coverage  & \bf diff.       \\ \hline
shunting			&20	 & 86\%	 &12	&46\%	&40\%  \\
dc\_1				&8	 &91\%	 &16	&89\%	&2\%  \\
dc\_2				&2	 &100\%	 &1	    &100\%	&0\%  \\
dc\_3				&6	 &100\%	 &5	    &100\%	&0\%  \\
dc\_4	  		    &2	 &92\%	 &2	    &86\%	&6\%  \\
dc\_5	    		&2	 &89\%	 &1	    &89\%	&0\%  \\
dc\_6	    		&2	 &90\%	 &3	    &84\%	&6\%  \\
dc\_7	    		&2	 &80\%	 &1	    &50\%	&30\%  \\
dc\_8	   			&3	 &83\%	 &1	    &50\%	&33\%  \\
dc\_9	    		&9	 &100\%	 &6	    &100\%	&0\%  \\
dc\_10	    		&9	 &93\%	 &6	    &93\%	&0\%  \\
dc\_11	    		&2	 &100\%	 &1	    &100\%	&0\%  \\
dc\_12	    		&3	 &72\%	 &3	    &60\%	&12\%  \\
dc\_13	    		&4	 &98\%	 &4	    &82\%	&16\%  \\
dc\_14	    		&2	 &82\%	 &1	    &64\%	&18\%  \\
radiohole 			&6	 &95\%	 &4	    &68\%	&27\%  \\
crossnonlx 			&13	 &84\%	 &6		&19 \%  &65\%   \\
baliseinfo 			&3	 &97\%	 &2	    &45\%	&52\%  \\  
emergency 1 		&14	 &94\%	 &1	    &6\%	&88\%  \\
emergency 2 	    &6	 &82\%	 &6	    &54\%	&28\%  \\
mema 				&7	 &89\%	 &5	    &41\%	&49\%  \\
trackside			&3	 &99\%	 &5	    &20\%	&79\%  \\
vbc 				&12	 &94\%	 &3	    &40\%	&54\%   \\
coordfromrbc 		&5	 &83\%	 &7	    &57\%	&26\%  \\
adfactordmi 1 		&3	 &85\%	 &2	    &71\%	&14\%  \\
adfactordmi 2 		&2	 &96\%	 &3	    &96\%	&0\%  \\
driveridins 		&10	 &89\%	 &2	    &65\%	&24\%  \\
eirene				&3	 &94\%	 &3	    &66\%	&28\%  \\
ertmslevel 			&3	 &94\%	 &4	    &88\%	&6\%  \\
natvalues 			&4	 &90\%	 &-		& -     &  -   \\
networkidins 		&3	 &94\%	 &2	    &75\%	&19\%   \\
rbcidins			&4	 &95\%	 &2	    &49\%	&46\%   \\
trainDataUpdate 	&1	 &89\%	 &4	    &60\%	&29\%   \\
trainDataInsertion 	&3	 &95\%	 &6	    &89\%	&6\%   \\
message129 			&10	 &83\%	 &8	    &77\%	&6\%    \\
runnumber 1			&3	 &94\%	 &2	    &70\%	&24\%   \\
runnumber 2 		&7	 &92\%	 &7	    &93\%	&-1\%   \\

\hline
\end{tabular}
\end{change3}
\end{table}

\begin{change3}
Table \ref{tab:AFL} reports on test cases generated with AFL for the programs considered in our case study (but program \texttt{natvalues} for which AFL unexpectedly generated a broken instrumentation that made the program crash deterministically at runtime). The table
indicates the information of the test cases generated with \tool when equipped with \klee (columns \textit{\tool}), in comparison with the number of test cases and corresponding model coverage data achieved with AFL (columns \textit{AFL}), and shows the difference between the coverage measurements in either case (column \textit{diff}).\end{change3}

The data in the table indicate that the two approaches led to generating test suites of comparable size in most cases, but the model coverage achieved with AFL was often significantly lower than the coverage achieved with \tool.
\begin{change3}
AFL achieved the same amount model coverage as \tool for 7 programs (namely, \texttt{dc\_2}, \texttt{dc\_3}, \texttt{dc\_5}, \texttt{dc\_9}, \texttt{dc\_10}, \texttt{dc\_11} and \texttt{adfactordmi 2}), achieved more coverage than \tool only for 1 program (namely, \texttt{runnumber 2}), and achieved less coverage than \tool for the remaining 28 programs. In the 28 cases in which \tool outperformed AFL, the difference in coverage ranged between 2\% and 88\%, with a median of 26\%. 
In the only case in which AFL outperformed \tool, the difference in coverage was rather limited (1\%) due to a single MC/DC objective that \tool did not cover because it 
missed a specific truth value for a condition that did not belong to the path condition of the corresponding execution path, while AFL could hit by mutating inputs at random.\end{change3}
We interpret these data as clear evidence that a tool like AFL does not suite for our goal of testing safety-critical software, and we believe that the type of weaknesses that we observed by experiencing with AFL likely generalize to search-based test generators at the state of the art.

\end{change2}

\subsection{Threats to validity}
\label{sec:threats}

The main internal threats to the validity of our findings are concerned with the risk of implementation errors in \tool (that could bias our results), and with use of coverage indicators to evaluate the quality of the test suites.

We extensively tested \tool to ascertain its correctness, and manually crosschecked  several result samples.
For the implementation of the symbolic execution phase, which is at the core of the results that we computed with \tool, we relied on \klee,  a state of the art symbolic executor actively maintained and largely used  in the community. Thus we are confident in the validity of the results that we obtained with \tool.

We evaluated the quality of the test suites that \tool computed in our experiments
based on the model coverage indicators obtained with the tool \scade Test.
We drew on the documentation provided from Ansys and on the
advising of the industrial partner with whom we are collaborating, to reckon that the
coverage indicators computed with \scade Test correspond to domain-relevant coverage requirements.
However, we are well aware that any coverage measurement is just a proxy of the effectiveness of the test cases, and we cannot take for granted that high coverage rates necessarily correspond to high fault-detection power. \begin{change3}
We attacked this issue by showing that 
the test suites generated by our tool, once complemented with assertion-style test oracles, succeeded in revealing component-level faults of the considered \scadelang programs.
\end{change3}
We look forward to experiencing \tool on further \scade programs that are currently being developed in the project, to collect further data on which faults we can indeed detect with the help of the test cases generated with \tool.

The external threats to validity relate to the extent to which our finding can generalize. So far, 
we experienced \tool against the set of subject programs considered in this paper, which are admittedly only a small sample of the possible safety-critical programs. 
Nonetheless, on one hand, these programs are a representative sample of the safety-critical software that our industrial partner typically develops,  
following the most prominent certification standards in the railway sector; On the other hand,
the restrictions that \scade embraces to promote the safety of the programs are common to other programming languages for developing safety critical software, e.g., \textsc{SaferC}. Thus, we believe that our result might in fact generalize.

 \section{Related Work}\label{sec:related}
We surveyed the most relevant techniques for automated test generation for software programs in the introduction of this paper, encompassing test generators based on
random testing, search-based testing and  symbolic execution. 
Random testing and search-based testing derive  test cases  by either randomly sampling the possible program inputs or based on dynamic data about the execution of the programs. Symbolic execution systematically unfolds the execution space of the programs under test and generates test cases by solving the execution of the possible program paths. 
Then we have then described our approach to test case generation for safety-critical programs in \scadelang, which is based on symbolic execution, and compared our approach with a analogous embodiment based on search-based testing by referring to the tool AFL.
\begin{change}
Other results on the effectiveness of automated test generation in safety critical systems are provided in~\cite{Enoiu2016,Gay2015}.\end{change}

The research that we described in the paper is also related to other approaches for automated model-based testing, to methods for formally specifying and verifying safety-critical software, and to other pieces of research on verifying programs in \scadelang.

\subsection{Automated model-based testing}

Our approach can be seen as related to 
\emph{model-based testing}, which derives test cases by analyzing program specifications or program behaviors expressed in suitable modelling languages, e.g., UML class diagrams, state machines or sequence diagrams~\cite{UttingLegeard2010,manual:uml}.
Model-based testing has been  successfully applied to complement verification of formal specifications expressed in languages as B, Z or VDM~\cite{hierons2015}. For a comprehensive survey of model-based testing we refer the readers to the work  of Utting et al.~\cite{UttingEtAl2012} and Dias Neto et al.~\cite{10.1145/1353673.1353681}.

The approach that we presented in this paper addresses the test generation problem based on the analysis of the execution paths in the programs, and naturally lends itself to complement or be complemented 
with further test cases generated either manually or yet automatically in model-based fashion.

\begin{change2} In particular our approach shares similarities with the ones of Polyglot ~\cite{polyglot,polyglot2} and SAUML \cite{sauml}, which exploit symbolic execution to generate test cases for systems modeled with statechars and UML-RT state machines, respectively. 
Polyglot is similar to \tool in that it translates statecharts to programs (specifically programs in Java) and then exploits symbolic execution (by means of the symbolic executor  SPF~\cite{SPF} that addresses Java),  to generate test cases that achieve path coverage  up to some specified depth. 
SAUML extends symbolic execution to directly analyze the UML-RT models (i.e., it works without converting the models to programs)
to check properties like reachability and invariants,
and to generate test cases. 
Our approach  differs from both these approaches in the way \tool distinctively uses symbolic execution within an analysis algorithm tailored on the characteristics of the \scadelang models, which foster programs with finite path spaces and input data structures comprised of finite sets of distinct fields.

\end{change2}

\subsection{Formal methods for safety-critical software}
Safety-critical systems need to strictly comply with their requirements as they were elicited in the earliest phases of the development process. \emph{Formal} methods~\cite{10.1145/1134285.1134406} define one or more languages with mathematically precise semantics that can be used to describe the requirements, the domain constraints and the designs, and to prove or disprove relevant properties thereby, e.g., absence of deadlock or unreachability of unsafe states. Most formal method define mathematically rigorous procedures to ensure that the artifacts produced at every step of a development process \emph{refine} the artifacts produced at earlier steps, thus preserving all their relevant properties. The downside of these approaches is the degree of mathematical sophistication that they demand to software engineers and designers, who should be able to model a system with a formal specification, prove (or disprove) its properties, refine an abstract (not directly computable) specification progressively to a concrete (computable) one, and translate a concrete specification to an executable program in a given programming language. To this end, formal methods are often accompanied with tools that assist in performing their tasks, with various degrees of automation, which anyway hardly balance the aforementioned complexity. 

Formal methods differ for the breadth of their scope. At one end of the spectrum, methods like B or its successor Event-B~\cite{Abrial2010} aim at producing a complete, correct-by-construction approach, encompassing all the phases of the development lifecycle. These methods usually refrain from testing the final implementation, in the assumption that having proved both a sufficient set of correctness properties on the abstract designs, and their preservation through the refinement steps may suffice to ensure that the final program is correct \emph{by-construction}. Other formal approaches do not have the generality of a full correct-by-construction method, and focus only on assisting a well defined part of the software development process. This is the case of Alloy~\cite{Jackson2012}, a language and a tool for modeling systems that is suited to assist the specification and abstract design activities. Similarly, Z~\cite{Spivey:ZNotation:1989} is customarily used as a system modeling language, although there also exists a well-established theory of refinement for Z~\cite{usingz}. 

Formal approaches that do not have the generality of correct-by-construction methods can benefit from software testing to provide some degree of assurance  that the derived implementations comply with the corresponding requirement specifications. Even correct-by-construction approaches might require  testing, to cope with the \emph{weak} (i.e., unproved) points of the refinement and translation chain,
or simply to comply with certification requirements~\cite{hierons2015}.

\subsection{Automated test generation for \scadelang models}
\scadelang can be regarded as a formal modeling approach focused on the detailed design and implementation phases of the software lifecycle. The \scadelang language is derived from the synchronous dataflow programming languages \lustre~\cite{10.1109/5.97300}, with some programming constructs derived from the programming language \esterel~\cite{10.1016/0167-6423(92)90005-V}  and from the graphical, state-machine-based language SyncCharts~\cite{andre1996}. \scadelang has formally defined semantics. All its constructs are computable, and therefore it is suited to express concrete designs rather than requirements and high-level system models. 
The \scade suite provides a model-based test coverage measurement tool that, from a \scadelang model and a test suite, calculates the coverage of different categories of elements in the model (states, transitions, conditions in transition guards, MC/DC coverage). 

To the best of our knowledge, the only research works that address automated test generation for \scadelang or \lustre programs is the work proposed in 
\cite{10.1109/ISSRE.2005.26}. This work introduces a set of coverage criteria for \lustre and \scadelang programs, defined over the graph of operators in the programs, and an automated tool that builds test suites that maximize these coverage criteria. The performance of the test generator is assessed by measuring the mutation analysis~\cite{10.1145/3155133.3155155}.

The above approach, however, differs in aim and scope from ours. Our approach systematically analyzes the C code that corresponds to the \scadelang programs, while the approach of~\cite{10.1109/ISSRE.2005.26} does not consider the generated C code. The authors of~\cite{10.1109/ISSRE.2005.26} propose dedicated coverage measures, specific for synchronous dataflow programming languages, while we aim at covering all execution paths in the programs.

\begin{change2}
An interesting tool 
is RT-Tester \cite{RTTester, RTTester2}, which is used in industry to perform V\&V activities for avionic, automotive and railway systems: it starts from a concrete test model describing the expected behaviour of
the system under test,  
renders the models into a set of expressions in propositional logic, and 
then solves the formulas 
with a SMT solver to generate test cases.
Bounded model checkers, like CBMC, take a similar approach~\cite{cbmc}. They represent programs with boolean formulas, which they then check for satisfiability by using a SAT solver, to generate test cases as counterexamples of verification properties.
In the future, we aim to compare with these approaches. 
\end{change2}

 \section{Conclusions}\label{sec:conclusions}

The development of safety-critical software must ensure with a high degree of confidence software programs that behave correctly in all operating conditions. To this end, automated software testing can assist in verifying the programs more thoroughly, more quickly, and at a lower cost than traditional, manual testing techniques. 
In this paper, we studied the viability of  
an automated test generation approach based on symbolic execution, \begin{change3}specifically tailored on the characteristics  
of a programming language for safety-critical software systems.\end{change3} We instantiated the proposed approach with the tool \tool, a test generator for programs written the \scadelang language. 
The case study that we reported in this paper indicates that the proposed approach 
was able 
to successfully produce test suites that
achieve a high model coverage \begin{change3}and assist in identifying faults\end{change3} 
for the considered safety-critical programs in \scadelang, 
while keeping the test generation effort under control.

We envision many opportunities for future research on the topic. We plan to extend the experimental assessment by considering further case studies.
On one hand, we aim at assessing the scalability of the proposed approach through the analysis of components with growing complexity. On the other hand, we would like to investigate the  
possibility of extending our approach to  safety-critical software developed in  programming languages other than \scadelang.
We also plan to extend the evaluation of \tool by assessing the fault detection ability of the generated test suites, e.g. by exploiting a mutation framework for \scadelang~\cite{10.1145/3155133.3155155}.

Lastly, the test cases generated by \tool currently contain assertion checks that are usable for regression testing only, but
in the future we would like to integrate \tool with a component for generating general oracles. 
Automatic oracle generation is an open research problem, and we are currently studying how to extend techniques to automatically generate oracles from software annotations~\cite{Goffi:Exceptional:ISSTA:2016} so that the oracles are generated from the software requirements specification documents.

\section*{Acknowledgement}
This work is partially supported by the SISMA national research project (MIUR, PRIN 2017, Contract 201752ENYB). 

We thank Raffaele Saponara for his contribution in running some of the experiments reported in the paper, our colleagues of the development team and our industrial partner for their help.

\appendix

\end{document}